\documentclass[]{jfm}

\usepackage{graphicx}
\usepackage{newtxtext}
\usepackage{newtxmath}
\usepackage{natbib}
\usepackage{hyperref}
\usepackage{subcaption,caption}
\usepackage{placeins}
\usepackage{comment}
\usepackage{bm}

\hypersetup{
	colorlinks = true,
	urlcolor   = blue,
	citecolor  = blue,
}

\newcommand{\RomanNumeralCaps}[1]
\linenumbers

\title{An extension of the compound flow theory with friction between the streams and at the wall}

\author{Jan Van den Berghe\aff{1,2}
	\corresp{\email{jan.vandenberghe@vki.ac.be}},
	Miguel A. Mendez\aff{1}
	\and Yann Bartosiewicz\aff{2}}

\affiliation{
	\aff{1} von Karman Institute for Fluid Dynamics, Waterloosesteenweg 72, 1640 Sint-Genesius-Rode, Belgium
	\aff{2} Institute of Mechanics, Materials, and Civil Engineering (iMMC), Université catholique de Louvain (UCLouvain),
	1348 Louvain-la-Neuve, Belgium
}

\begin{document}
	\newcommand\Ma{\mbox{{Ma}}}  
	
	\maketitle
	
	\begin{abstract}
		Compound flows consist of two or more parallel compressible streams in a duct and their theoretical treatment has gained attention for the analysis and modelling of ejectors. Recent works have shown that these flows can experience choking upstream of the geometric throat. While it is well known that friction can push the sonic section downstream the throat, no mechanism has been identified yet to explain its displacement in the opposite direction. This study extends the existing compound flow theory and proposes a 1D model, including friction between the streams and the duct walls. The model captures the upstream and downstream displacements of the sonic section. Through an analytical investigation of the singularity at the sonic section, it is demonstrated that friction between the streams is the primary driver of upstream displacement. The 1D formulation is validated against axisymmetric Reynolds Averaged Navier-Stokes (RANS) simulations of a compound nozzle for various inlet pressure and geometries. The effect of friction is investigated using an inviscid simulation for the isentropic case and viscous simulations with both slip and no-slip conditions at the wall. The proposed extension accurately captures the displacement of the sonic section, offering a new tool for in-depth analysis and modeling of internal compound flows.
		
	\end{abstract}
	
	
	
	\section{Introduction}
	Parallel compressible streams are at the heart of many applications, the most classic example being jet engines with a high-speed core and a slower surrounding jet. Examples of internal flows with multiple streams include ejectors or (sc)ramjets. Such heterogeneous flows are referred to as compound flows. This study investigates the choking of internal compound flows.
	
	Recently, \cite{LAMBERTS2018_compound} observed that compound flows could choke \emph{upstream} of the geometrical throat, in contrast to the classic \emph{downstream} displacement of the sonic section due to friction. By numerically increasing the wall roughness, \cite{kracik2023effect} showed that friction has this effect on compound flows as well. \cite{LAMBERTS2018_compound} attributed the upstream shift to non-isentropic effects, but the inherent cause remained unknown. While both \cite{LAMBERTS2018_compound} and \cite{kracik2023effect} addressed this problem through axisymmetric Reynolds-averaged Navier-Stokes (RANS) simulations, to the best of the authors' knowledge, no theoretical analysis of the sonic section displacement has been provided.
	
	The first theoretical treatment of compound flows was proposed by \cite{pearson1958theory}. These authors introduced the idea of splitting the domain into parallel 1D streams bounded by dividing streamlines and walls. The position of these streamlines can be identified by solving mass, momentum and energy conservation in each stream, together with a force balance at the dividing streamlines. The authors assumed adiabatic and isentropic conditions for the compound flow. Hence, only a pressure force is exerted between the streams, and the force balance requires equal static pressure for the entire cross-section. This formulation leads to a closed system of equations linking the evolution of mass, momentum and energy in each stream to the evolution of the respective cross-sections and streamwise pressure distribution. \cite{pearson1958theory} state that the assumption of equal pressure is reasonable if the curvature of the streamlines remains limited. 
	
	Through some manipulations of the conservation equations, \cite{bernstein1967compound} showed that a single equation can be obtained for the uniform static pressure. Similarly to a single 1D stream (see, for example, \cite{shapiro1953dynamics}), this equation is singular with a division of zero by zero if the flow is choked. The major result of the study by \cite{pearson1958theory} lies in the fact that a compound flow can be choked with a combination of subsonic and supersonic streams. They showed that information cannot travel upstream in the subsonic streams due to the assumption of uniform static pressure. The derivation by \cite{bernstein1967compound} confirms the findings of \cite{pearson1958theory} identifying the compound choking from the singularity of the pressure equation. The same compound choking condition was found by \cite{hoge1965choked} by minimising the impulse function defined in chapter 4 of \cite{shapiro1953dynamics}. An alternative approach by \cite{agnone1971comment} identifies the compound choking from the singularity of the governing equations, written in a matrix form, while \cite{fage1976apparent} defines it through the so-called `elasticity of the fluid boundary' between the streams. \cite{bernstein1967compound} gave the name `compound flow theory', which we also adopt. These works did not consider exchanges between the streams; hence, the flow remained isentropic.
	
	Later work investigated the effect of forces on the compound flow. \cite{hoge1965choked} explored the effects of friction between the streams and the walls, assuming isentropic and adiabatic conditions but without deriving an explicit model to account for their impact on the flow. These authors postulated that the exchange of momentum between the streams would have a minimal impact on the compound choking mechanism. A more explicit approach is presented by \cite{clark1995application}, \cite{papamoschou1996analysis} and \cite{grazzini2015constructal}, who used the correlation proposed by \cite{papamoschou1993model} to compute the shear stress between the streams. The authors used the resulting compound flow equations \emph{with} momentum exchange in a 1D ejector model but did not investigate the choking mechanism. Earlier work by \cite{otis1976choking} treats the mixing and choking of a compound flow for constant-area ducts. An important conclusion of that work is that additional effects such as forces or heat exchange do not affect the choking condition as derived in the original compound theory.
	To the best of the author's knowledge, a general treatment of compound choking with momentum exchange in variable area ducts has not yet been proposed in the literature.
	
	This work proposes such an extension and investigates the effect of friction between the streams and the walls on the choking mechanism. It is analytically shown that the sonic section moves upstream or downstream depending on the relative magnitude of the inter-stream and wall friction. The analysis is similar to `generalised 1D flow' \cite[chapter 8]{shapiro1953dynamics}, which describes quasi-1D compressible flow of a \emph{single} stream with any combination of area change, friction, heat exchange and/or mass injection. Similarly, the isentropic \emph{compound} flow theory, which only includes area change, is generalised in this work by including momentum exchanges. Our contribution is both theoretical and numerical: (1) we propose a simple theoretical model that captures the position of the sonic section, and (2) we propose a quasi-third-order approximation of the pressure gradient in the vicinity of the sonic section to circumvent the numerical issues near the singular sonic point.
	
	The governing equations are derived in Section \ref{sec:derivation} and converted into an explicit system of Ordinary Differential Equations (ODEs). Section \ref{sec:approx} goes into more detail on the singularity of the equation for the static pressure at the sonic section. The numerical solution of the model is discussed in Section \ref{sec:numerical_solution}. Numerical test cases are presented in Section \ref{sec:numerics}, to compare the results with the 1D predictions in Section \ref{sec:results}. Section \ref{sec:conclusion} closes the paper with conclusions and perspectives.

	\section{Model definition} \label{sec:derivation}
	The governing equations are derived in their general form in Section \ref{sec:derivation_governing}. Section \ref{sec:closure} introduces the treatment of the friction forces in the equation governing the streamwise pressure gradient. The corresponding choking mechanism is analysed in Section \ref{sec:choking}.
	
	\subsection{Governing equations} \label{sec:derivation_governing}
\begin{figure}
		\centerline{\includegraphics[width=\linewidth]{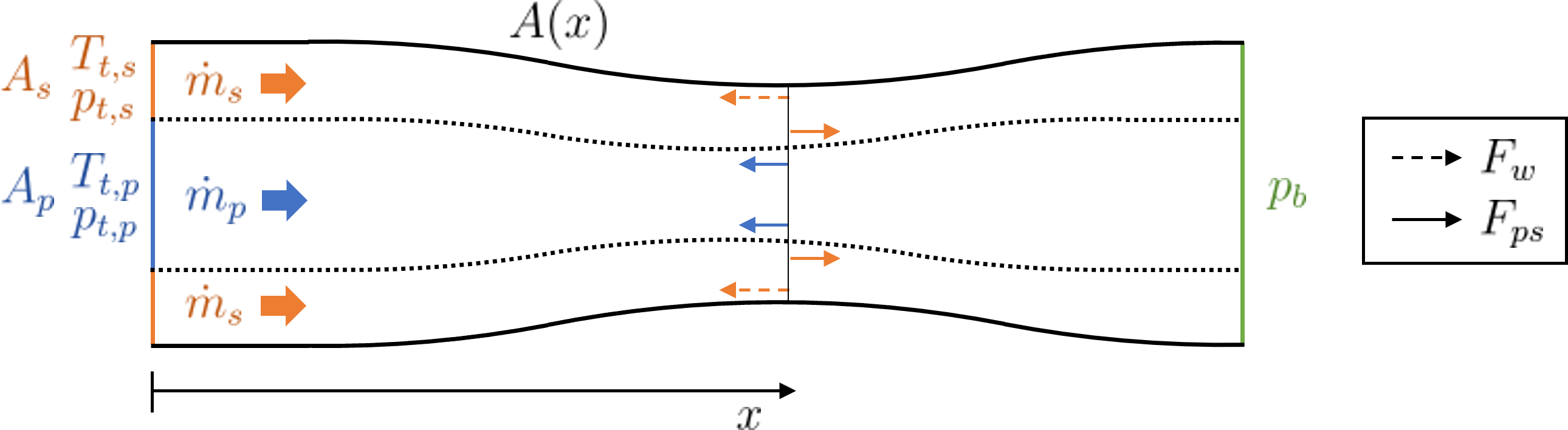}}
		\caption{The considered axisymmetric nozzle has a contour $A(x)$ and contains two streams, which have the same total temperature $T_t$, but can have a different total pressure $p_t$. They exchange momentum through a friction force $F_{ps}$ at the dividing streamline and a wall friction force $F_w$. The static pressure $p$ is assumed to be uniform in each cross-section. The solution consists of the streamwise distribution of the flow variables $p$, $p_{t,p}$, $p_{t,s}$, $T_{t,p}$, $T_{t,s}$ and the individual cross-sections $A_p$ and $A_s$.}
		\label{fig:sketch}
	\end{figure}

	Consider an internal flow consisting of two parallel streams (cf. figure \ref{fig:sketch}) of the same ideal gas. The theory can be extended to any number of streams, but this generalisation is left for future work. 
	The streams are modelled as uniform 1D streams, i.e., in terms of cross-stream averaged quantities. The main assumption for compound-compressible flows is that the static pressure is uniform in all cross-sections, hence:
	\begin{equation}
		p_p(x) = p_s(x) = p(x)\,, \label{eq:assump_p}
	\end{equation}
	where $p$ denotes the static pressure, $x$ is the axial coordinate and the subscripts $p$ and $s$ refer to the primary and secondary stream. The distribution of the cross-sections $A_i(x)$, with $i \in [p,s]$, is initially unknown and should be calculated. These are subject to a geometrical constraint imposed by the cross-section of the channel $A(x)$:
	\begin{equation}
		\label{sum_A}
		\sum_{i\in[p,s]} A_i(x) = A(x)\,.
	\end{equation}
	The governing equations consist of the conservation of mass, momentum and energy in each stream. For a steady, quasi-1D flow, these read as follows:
	\begin{align}
		\frac{d}{dx} (\rho_i A_i u_i) &= 0\,,\\
		\frac{d}{dx} (\rho_i A_i u_i^2) &= -A_i \frac{dp_i}{dx} + F_i\,, \label{eq:momentum}\\
		\frac{d}{dx} (\rho_i A_i u_i h_{t,i}) &= 0\,,
	\end{align}
	where $\rho$ denotes the density, $u$ denotes the axial velocity, $F_i$ is the net force per unit length exerted on stream $i$ and $h_t$ denotes the total enthalpy. The force $F_i$ accounts for the friction between the streams and at the wall, and is defined in Section \ref{sec:closure}. The equations imply that the streams do not exchange mass and that the separating line between the streams is the \emph{dividing streamline}. Heat exchange is not considered.\\
	This system of equations can be converted to an explicit system of ODEs for numerical integration. The derivation can be found in classic textbooks; for example in chapter 8 of \cite{shapiro1953dynamics}. The equations in terms of the static pressure $p$, the total pressure $p_t$ and the total temperature $T_t$ read as follows:
	\begin{align}
		\frac{1}{p_i} \frac{dp_i}{dx} &=  \left[\frac{1 + \left(\gamma  - 1\right)\Ma_i^2}{1-\Ma_i^2}\right] \frac{F_i}{A_i p_i} + \left[\frac{\gamma \Ma_i^2}{1-\Ma_i^2}\right]\frac{1}{A_i}\frac{d A_i}{dx}\,, \label{eq:p_i}\\
		\frac{1}{p_{t,i}}\frac{dp_{t,i}}{dx} &= \frac{F_i}{A_i p_i}\,,\label{eq:pt_i}\\
		\frac{1}{T_{t,i}}\frac{dT_{t,i}}{dx} &= 0\,,\label{eq:Tt_i}
	\end{align}
	having introduced the ratio of specific heats $\gamma$ and the Mach numbers $\Ma_i$:
	\begin{align}
		\Ma_i =  u_i / a_i\,, \quad \mbox{and } \quad a_i = \sqrt{\gamma R T_i} \, ,
	\end{align}
	where $R$ denotes the specific gas constant. Equations \eqref{eq:p_i}-\eqref{eq:Tt_i} do not form a closed system of equations because the cross-sections $A_i(x)$ are unknown (as opposed to a single stream where $A_i = A$). Nonetheless, their sum should equal the known channel's cross-section $A(x)$ \eqref{sum_A}. The derivative $dA/dx$ can be computed from the channel's profile, so the unknown terms $dA_i/dx$ can be eliminated from the system by summing over the streams $i$. The gradient of the cross-section $A_i$ is first isolated from equation \eqref{eq:p_i}:
	\begin{equation}
		\frac{dA_i}{dx} = \left[A_i\frac{1-\Ma_i^2}{\gamma \Ma_i^2}\right]\frac{1}{p} \frac{dp}{dx} - \left[\frac{1 + \left(\gamma  - 1\right)\Ma_i^2}{\gamma \Ma_i^2}\right] \frac{F_i}{p}\,, \label{eq:A_i}
	\end{equation}
	where the index for the pressure is omitted because of \eqref{eq:assump_p}. Note that \eqref{eq:A_i} can be integrated to compute the cross-section of a single stream if the pressure gradient is known. Summing over the streams $i$ gives:
	\begin{equation}
		\sum_{i\in[p,s]} \frac{dA_i}{dx} = \frac{dA}{dx} = \sum_{i\in[p,s]} \left[A_i\frac{1-\Ma_i^2}{\gamma \Ma_i^2}\right]\frac{1}{p} \frac{dp}{dx} - \sum_{i\in[p,s]}\left[\frac{1 + \left(\gamma  - 1\right)\Ma_i^2}{\gamma \Ma_i^2}\right] \frac{F_i}{p}\,. \label{eq:sum_Ai}
	\end{equation}
	The pressure gradient is the only remaining unknown, for which an explicit equation is obtained from \eqref{eq:sum_Ai}:
	\begin{equation}
		\frac{1}{p} \frac{dp}{dx} = \dfrac{1}{\beta}\left( \dfrac{dA}{dx} + \sum\limits_{i\in[p,s]}\left[\dfrac{1 + \left(\gamma  - 1\right)\Ma_i^2}{\gamma \Ma_i^2}\right] \dfrac{F_i}{p}\right)\,, \label{eq:p_compound}
	\end{equation}
	where $\beta$ denotes the compound choking indicator:
	\begin{equation} \label{eq:beta}
		\beta = \sum_{i\in[p,s]} A_i\dfrac{1-\Ma_i^2}{\gamma \Ma_i^2}\,.
	\end{equation}
	
	\cite{bernstein1967compound} showed that $\beta > 0$ corresponds to a compound-subsonic flow, $\beta=0$ to a compound-sonic flow and $\beta < 0$ to a compound-supersonic flow. As a result, the pressure gradient and the area gradient of a compound-subsonic flow share the same sign in the absence of forces, leading to an expansion in a convergent channel and a compression in a divergent channel. Recently, \cite{METSUE2021121856} extended the compound choking condition to real gases and showed that the condition $\beta=0$ corresponds to the maximal combined mass flow rate.
	
	A convenient variable to identify the compound flow regime is the so-called equivalent Mach number $\Ma_{eq}$, introduced by \cite{hedges1974compressible}:
	\begin{equation}
		\Ma_{eq} = \left(\gamma \frac{\beta}{A}+1\right)^{-\frac{1}{2}}\,. \label{eq:Ma_eq}
	\end{equation}
	
	This equals one for a compound choked flow. Section \ref{sec:choking} analyses choking in more detail. It is important to note that the forces do not affect $\beta$ and hence the compound choking criterion, as pointed out by \cite{otis1976choking}. Moreover, as shown by \cite{bernstein1967compound}, compound waves can no longer travel upstream a location $x$ if $\beta(x) \leq 0$, which constitutes the fundamental principle of choking as in the case of a single stream.
	
	Equations \eqref{eq:pt_i}, \eqref{eq:Tt_i}, \eqref{eq:A_i} and \eqref{eq:p_compound} form a closed system for the (uniform) static pressure $p$ and the total pressure $p_{t,i}$, the total temperature $T_{t,i}$ and the cross-section $A_i$ in each stream. These variables fully define the state of the flow and other variables can easily be calculated from them using classic gas dynamic relations. The explicit nature of the equations facilitates numerical integration with a shooting method. More details on the numerical solution are provided in Section \ref{sec:numerical_solution}.
	
	\subsection{Model closure} \label{sec:closure}
	The compound flow is assumed to be axisymmetric, with the primary stream surrounded by the secondary stream. Planar flows are left as technical extensions for future studies. Hence, wall friction acts only on the secondary stream. A friction coefficient $f_w$ is defined such that the friction force per unit length $F_w$ equals:
	\begin{equation}
		F_w = \frac{1}{2} f_w \rho_s u_s^2 l_w = \frac{1}{2} f_w \gamma p \Ma_s^2 l_w\,, \label{eq:F_w}
	\end{equation}
	where $l_w$ denotes the wetted perimeter of the cross-section. In a circular duct, $l_w$ is related to the cross-section $A$ as follows:
	\begin{equation}
		l_w = 2 \pi r = 2 \sqrt{\pi A}\,, \label{eq:l_w}
	\end{equation}
	where $r$ denotes the radius of the total cross-section $A$. 
	Similarly, the friction between the streams is defined as \cite{papamoschou1993model}:
	\begin{equation} \label{eq:stream_friction}
		F_{ps} = \frac{1}{2} f_{ps} \dfrac{\rho_p+\rho_s}{2} \left(u_p - u_s\right) \left|u_p - u_s\right| l_{ps}\,,
	\end{equation}
	where $f_{ps}$ denotes a friction coefficient and $l_{ps}$ denotes the perimeter of the primary stream:
	\begin{equation}
		l_{ps} = 2 \pi r_p = 2 \sqrt{\pi A_p}\,. \label{eq:l_ps}
	\end{equation}
	
	The friction force between the streams opposes the stream with the highest velocity and is defined to be positive by \eqref{eq:stream_friction} if $u_p > u_s$. Therefore, $F_{ps}$ opposes the primary stream, while the secondary stream is entrained by the inter-stream friction and decelerated by the wall friction:
	\begin{equation}
		F_p = - F_{ps}\,, \quad \mbox{and } \quad F_s = F_{ps} - F_w\,. \label{eq:F_i}
	\end{equation}
	
	Two approaches were used in this work for both coefficients $f_w$ and $f_{ps}$ and tested to the RANS simulations in Section \ref{sec:results_closure}.
	
	The first approach was to leverage known correlations for similar flow configurations. \cite{vandriest1951turbulent}'s formula for the local wall friction coefficient $f_w$ under a turbulent boundary layer in a compressible flow gives:
	\begin{equation}
		\dfrac{0.242}{\sqrt{f_w}} \sqrt{1 - \lambda^2} \dfrac{\arcsin (\lambda)}{\lambda} = 0.41 + \log_{10} \left(f_w \Rey_x\right) + \log_{10} \left((1 - \lambda^2)\left(1 - \dfrac{\theta \lambda^2}{1 + \theta}\right)\right)\,,\label{eq:f_w_correlation}
	\end{equation}
	where
	\begin{equation}
		1-\lambda^2 = \left(1 + \frac{\gamma - 1}{2} \Ma_s^2\right)^{-1}\,, \quad \Rey_x = \dfrac{\rho_s u_s x}{\mu_s}\,, \quad  \, \quad \mbox{and } \quad \theta = \dfrac{S}{T_s}\,, 
	\end{equation}
	with the dynamic viscosity $\mu$ computed with \cite{sutherland1893lii}'s law, a reference viscosity $\mu_{ref}$ of $1.716 \ 10^{-5}$ Pa s at $T_{ref} = 273.2$ K, and $S = 110.4$ K.
	For the inter-stream friction coefficient $f_{ps}$, \cite{papamoschou1993model}'s formula gives
	\begin{equation}
		f_{ps} = 0.013 \dfrac{(1 + u_s/u_p) (1 + \sqrt{\rho_s / \rho_p})}{1 + (u_s/u_p) \sqrt{\rho_s / \rho_p}} \left(0.25 + 0.75 \exp(-3 \Ma_c^2)\right)\,, \label{eq:f_ps_correlation}
	\end{equation}
	with all quantities evaluated locally and the local convective Mach number $\Ma_c$ is defined as follows:
	\begin{equation}
		\Ma_c = \dfrac{u_p - u_s}{a_p + a_s}\,.
	\end{equation}
	
	The second approach was to infer these coefficients from the RANS results using an inverse method. In this setting, both coefficients $f_w$ and $f_{ps}$ are assumed to be constant and are determined by an optimizer which seeks to minimize the $l_2$ norm of the discrepancy between the proposed model and the results of RANS simulations. This approach is expected to be more accurate due to the calibration, provided that the assumption of constant coefficient is adequate. On the other hand, these coefficients are not expected to generalize well outside the reference data, while the correlations \eqref{eq:f_w_correlation} and \eqref{eq:f_ps_correlation} make the model self-sufficient and closed.
	
	\subsection{Choking mechanism}
	\label{sec:choking}
	Substituting the forces $F_p$ and $F_s$ as defined in the section above in equation \eqref{eq:p_compound} leads to a rational equation for the static pressure gradient:
	
	\begin{equation}
		\frac{1}{p} \frac{dp}{dx} = \dfrac{N}{\beta}\,, \label{eq:p_compound_2stream}
	\end{equation}
	with $\beta$ defined in \eqref{eq:beta} and the numerator:
	\begin{equation} \label{eq:numerator}
		N = \dfrac{dA}{dx} + N_{ps} + N_w\,,
	\end{equation}
	where 
	\begin{align}
		N_{ps} &= \dfrac{f_{ps}l_{ps}}{2}\frac{\Ma_p^2 - \Ma_s^2}{\gamma \Ma_p^2 \Ma_s^2}  \dfrac{\rho_p+\rho_s}{2} \left(u_p - u_s\right) \left|u_p - u_s\right| \dfrac{1}{p}\,,\\
		N_w &=  - \dfrac{f_{w} l_{w}}{2}  \left(1 + \left(\gamma  - 1\right)\Ma_s^2\right)\,.
	\end{align}
	
	The term $N_w$ stems from wall friction ($f_w$) and is negative by definition. The term $N_{ps}$ is due to friction between the streams and it is positive unless the terms $\Ma_p^2-\Ma_s^2$ and $u_p-u_s$ have a different sign. In the natural case $u_p>u_s$, this is possible only if $T_p/T_s \geq u^2_p/u^2_s$. This scenario is not encountered in \cite{LAMBERTS2018_compound} and \cite{kracik2023effect} and thus not discussed further. Therefore, the inter-stream friction has a positive contribution to the numerator $N$ in \eqref{eq:p_compound_2stream} in the considered conditions and the wall friction has a negative contribution. This leads to some fundamental implications.  
	
	A compound-subsonic flow ($\beta > 0$) expands in a convergent section ($dA/dx < 0$) and under the action of wall friction (as in a classic Fanno flow). Conversely, it recovers pressure in a divergent section and under the action of inter-stream friction. This force opposes the stream with the highest velocity, seeking to equalize velocities between streams. Consequently, momentum exchange between the streams promotes flow uniformity and results in a net increase in static pressure. The opposite is true for a compound-supersonic flow. 
	
	Equation \eqref{eq:p_compound_2stream} is singular in the sonic section ($\beta = 0$), similarly to a single stream where the denominator equals $1 - \Ma^2$ (cf. \cite{shapiro1953dynamics, restrepo2022viscous}). Nevertheless, classic nozzle flows and compound flows expand continuously through the sonic section, meaning that their pressure gradient is finite at $\beta=0$. This is possible according to equation \eqref{eq:p_compound_2stream} only if both the numerator $N$ and the denominator $\beta$ tend to zero, leading to an indeterminate form with a finite value. This observation is corroborated by \eqref{eq:sum_Ai}, which simplifies to $N=0$ if $\beta=0$. In isentropic conditions ($f_{ps} = f_w = 0$), a compound flow becomes sonic at the throat ($dA/dx = 0$), as shown by \cite{bernstein1967compound}. In non-isentropic conditions, the relative magnitudes of $N_{ps}$ and $N_w$ in \eqref{eq:numerator} determine the position of the sonic section.
	
	Wall friction has a negative contribution to the numerator. Therefore, in the absence of inter-stream friction ($f_{ps}=0$), one must have $dA/dx>0$ to respect $N=0$. This implies that the sonic section is located in a divergent section, hence downstream of the throat. The wall friction has the same effect on a nozzle flow with a single stream \citep{shapiro1953dynamics, beans1970computer, restrepo2022viscous}. Likewise, in the absence of wall friction ($f_w=0$), one must have $dA/dx < 0$ to respect $N=0$. This implies that the inter-stream friction pushes the sonic section upstream of the throat. This analysis offers a plausible explanation for the observed displacement of the sonic section by \cite{LAMBERTS2018_compound} and \cite{kracik2023effect}. The interplay between both forces is investigated further in Section \ref{sec:results}.
	
	The indeterminate pressure gradient can be computed by applying de l'Hôpital's rule on equation \eqref{eq:p_compound_2stream} (cf. \cite{bernstein1967compound, restrepo2022viscous}):
	\begin{equation}
		\left(\frac{1}{p}\frac{dp}{dx}\right)^* = \frac{N^*}{\beta^*} = \frac{0}{0} = \frac{(dN/dx)^*}{(d\beta / dx)^*}, \label{eq:p_delhopital}
	\end{equation}
	where the superscript $*$ refers to the sonic section. The derivatives $dN/dx$ and $d\beta/dx$ give rise to the derivatives of Mach numbers $\Ma_i$, cross-sections $A_i$, forces $F_i$ and the static pressure $p$ (see \eqref{eq:numerator} and \eqref{eq:beta}), and these can be all related to the local pressure gradient. The relevant equations are presented in appendix \ref{sec:dp_dx_first_order}.
	
	For example, the gradient of $\beta$ is given by:
	\begin{equation}\label{eq:dbeta_dx}
		\dfrac{d\beta}{dx} = \sum_{i\in[p,s]} \left(\dfrac{dA_i}{dx}\dfrac{1-\Ma_i^2}{\gamma \Ma_i^2} - \dfrac{A_i}{\gamma \Ma_i^4}\dfrac{d\Ma_i^2}{dx}\right)\,.
	\end{equation}
	
	Writing the Mach numbers $\Ma_i$ as a function of the total and the static pressure:
	\begin{equation} \label{eq:Mach_f_pt_p}
		\Ma_i^2 = \frac{2}{\gamma-1}\left(\left(\frac{p_{t,i}}{p}\right)^\frac{\gamma-1}{\gamma}-1\right)\,,
	\end{equation}
	the derivatives $d\Ma_i^2/dx$ in \eqref{eq:dbeta_dx} can be substituted by the gradients of the static and total pressures.
	The resulting expression for the gradient of $\beta$ thus contains first derivatives of the static pressure, the cross-sections $A_i$ and the total pressures $p_{t,i}$. The latter two can be substituted using \eqref{eq:pt_i} and \eqref{eq:A_i}, which contain the static pressure gradient. The resulting expression for the gradient of $\beta$ is linear with respect to the static pressure gradient, with non-linear coefficients depending on the Mach numbers, forces and static pressure:
	\begin{equation}
		\dfrac{d\beta}{dx} = c_{\beta0}(\Ma_p,\Ma_s)\bigg(\dfrac{1}{p}\dfrac{dp}{dx}\bigg) + c_{\beta1}(\Ma_p,\Ma_s,F_p,F_s,p)\,.
	\end{equation}
	
	The analytical expressions of $c_{\beta0}$ and $c_{\beta1}$ are given by \eqref{eq:cbeta0} and \eqref{eq:cbeta1}. The same approach leads to a similar linear equation for the derivative of the numerator (see \eqref{eq:dN_dx_final}). These equations are derived for constant friction coefficients, i.e., the correlations \eqref{eq:f_w_correlation} and \eqref{eq:f_ps_correlation} are not explicitly differentiated. This approximation is deemed acceptable since the distribution of $f_w$ and $f_{ps}$ are relatively flat, as later shown in Section \ref{sec:results}. The result is an implicit equation for the static pressure gradient in the sonic section:
	\begin{equation}
		\left(\frac{1}{p}\frac{dp}{dx}\right)^* = \dfrac{c_{N0}^* \bigg(\dfrac{1}{p}\dfrac{dp}{dx}\bigg)^* + c_{N1}^*}{c_{\beta0}^*\bigg(\dfrac{1}{p}\dfrac{dp}{dx}\bigg)^* + c_{\beta1}^*}. \label{eq:p_quadratic}
	\end{equation}
	
	Rearranging the equation above leads to the quadratic equation \eqref{eq:app_p_quadratic} for the static pressure gradient. The two roots correspond to the subsonic and the supersonic solution downstream of the sonic point (cf. \cite{shapiro1953dynamics}). For an isentropic compound flow, the derivative of the numerator corresponds to the second derivative of the cross-section $d^2A/dx^2$ ($f_{ps}=f_w=0$), leading to the tractable expression reported by \cite{bernstein1967compound}. The additional terms introduced by the friction forces preclude such a concise expression, which is why the coefficients and their analytical expressions are reported in the appendix \ref{sec:dp_dx_first_order}. Note, however, that these equations are exact and analytical.
	
	\vspace{-2mm}
	\section{Approximations to circumvent the singularity at the sonic section} \label{sec:approx}
	The pressure gradient can be computed in any section using \eqref{eq:p_quadratic} if the flow is compound-sonic or by using \eqref{eq:p_compound_2stream} otherwise. However, equation \eqref{eq:p_compound_2stream} is numerically ill-conditioned near the sonic section. Therefore, approximations of the pressure gradient near the sonic section help stabilising the numerical integration. In this section, such approximations are developed analytically.
	
	De l'Hôpital's rule used in Section \ref{sec:choking} arises from a Taylor series expansion of the numerator and the denominator around the sonic section $x^*$, which is then evaluated at the sonic section $x^*$ itself. More generally, these expansions lead to the following equation in $x$, which is still exact:
	\begin{equation}
		\frac{1}{p}\frac{dp}{dx} = \frac{N}{\beta} = \frac{N^* + (x-x^*) \frac{dN}{dx}^* + \frac{1}{2} (x-x^*)^2 \frac{d^2N}{dx^2}^* + \frac{1}{6} (x-x^*)^3 \frac{d^3N}{dx^3}^* + ...}{\beta^* + (x-x^*) \frac{d\beta}{dx}^* + \frac{1}{2} (x-x^*)^2 \frac{d^2\beta}{dx^2}^* + \frac{1}{6} (x-x^*)^3 \frac{d^3\beta}{dx^3}^* + ...}\,. 
	\end{equation}
	
	Since $N^* = \beta^* = 0$, the equation above simplifies to the following expression:
	\begin{equation}
		\frac{1}{p}\frac{dp}{dx} = \frac{N}{\beta} = \frac{\frac{dN}{dx}^* + \frac{1}{2} (x-x^*) \frac{d^2N}{dx^2}^* + \frac{1}{6} (x-x^*)^2 \frac{d^3N}{dx^3}^* + ...}{\frac{d\beta}{dx}^* + \frac{1}{2} (x-x^*) \frac{d^2\beta}{dx^2}^* + \frac{1}{6} (x-x^*)^2 \frac{d^3\beta}{dx^3}^* + ...}\,. \label{eq:p_taylor}
	\end{equation}
	
	Truncating the Taylor series leads to approximations of arbitrary order in space. The first order approximation only retains the first derivatives, which are constant since they are evaluated at the sonic section. Hence, the pressure gradient in the vicinity of the sonic section is approximated as a constant, equal to its value in the sonic section. Note that the resulting expression is identical to \eqref{eq:p_delhopital} and thus leads to the quadratic \eqref{eq:p_quadratic} in Section \ref{sec:choking}.

	\begin{figure}
		\center
		\begin{subfigure}[t]{0.48\linewidth}
			\includegraphics[width=\linewidth]{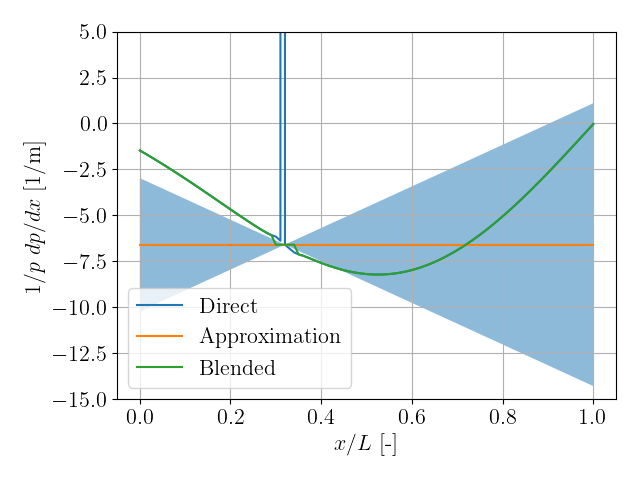}
			\subcaption{First order}
			\label{fig:dpdx_first}
		\end{subfigure}
		\begin{subfigure}[t]{0.48\linewidth}
			\includegraphics[width=\linewidth]{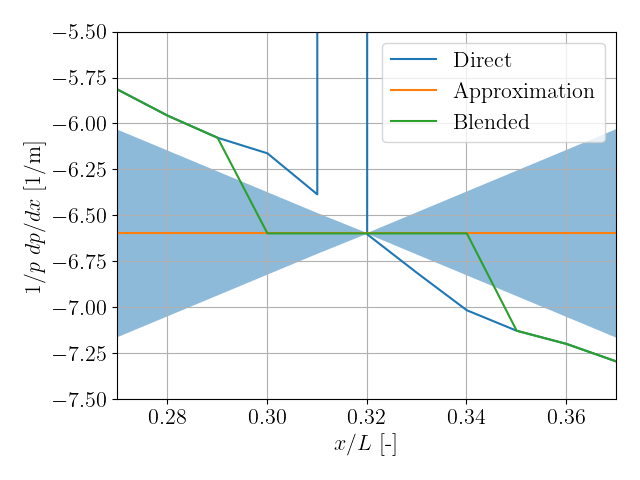}
			\subcaption{First order (zoom)}
			\label{fig:dpdx_zoom_first}
		\end{subfigure}
		\\
		\begin{subfigure}[t]{0.48\linewidth}
			\includegraphics[width=\linewidth]{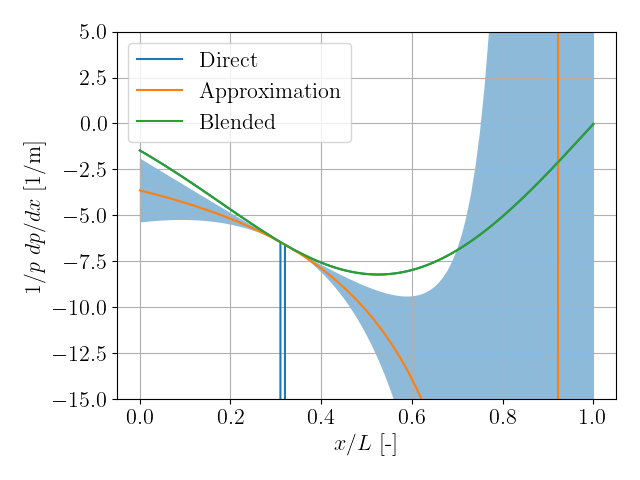}
			\subcaption{Second order}
			\label{fig:dpdx_second}
		\end{subfigure}
		\begin{subfigure}[t]{0.48\linewidth}
			\includegraphics[width=\linewidth]{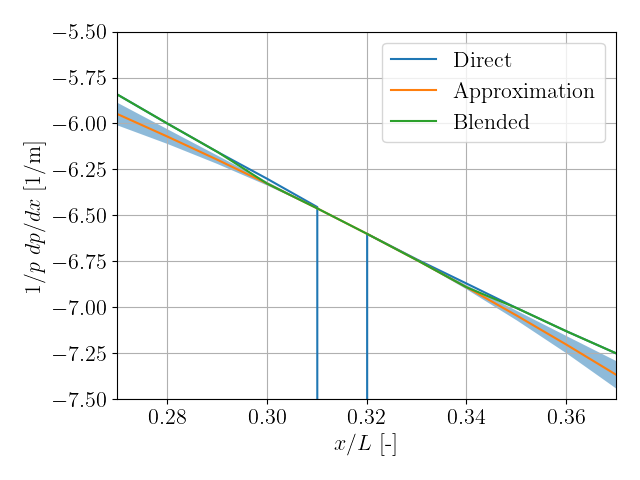}
			\subcaption{Second order (zoom)}
			\label{fig:dpdx_zoom_second}
		\end{subfigure}
		\\
		\begin{subfigure}[t]{0.48\linewidth}
			\includegraphics[width=\linewidth]{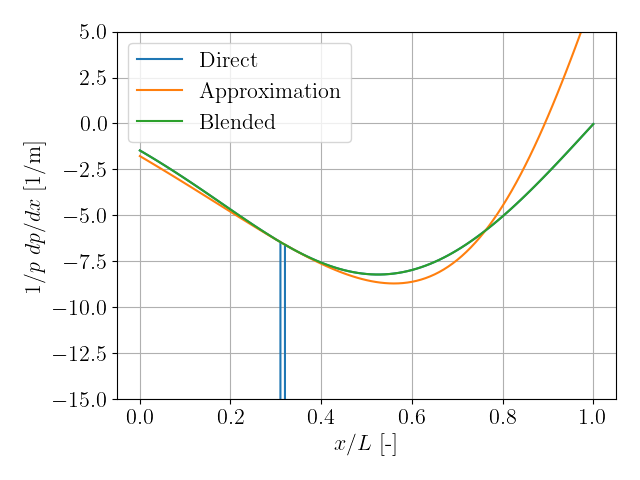}
			\subcaption{Quasi-third order}
			\label{fig:dpdx_third}
		\end{subfigure}
		\begin{subfigure}[t]{0.48\linewidth}
			\includegraphics[width=\linewidth]{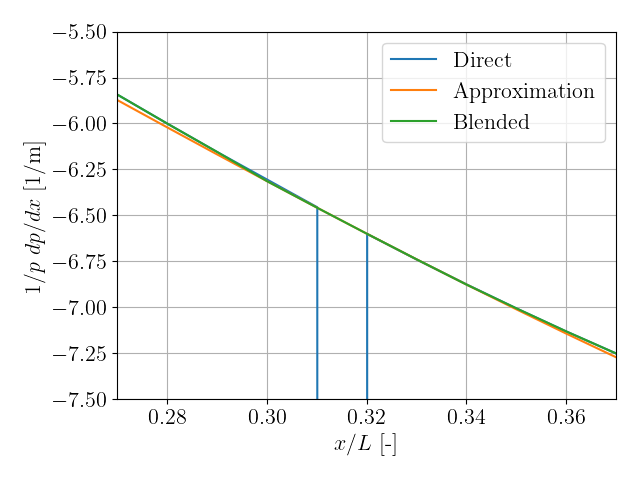}
			\subcaption{Quasi-third order (zoom)}
			\label{fig:dpdx_zoom_third}
		\end{subfigure}
		\caption{The pressure gradient computed directly with \eqref{eq:p_compound_2stream}, with the approximation \eqref{eq:p_taylor} (truncated after the first, second and third derivatives) and with the blended function \eqref{eq:dp_dx_blended}. The shaded region indicates the truncation error estimated by \eqref{eq:uncert}, which is unavailable for the quasi-third order approximation.}
		\label{fig:dpdx}
	\end{figure}
	
	Higher accuracy is obtained by retaining higher order derivatives. The analytical expressions for the second derivatives are given by \eqref{eq:d2N_dx2_final} and \eqref{eq:d2beta_dx2_final}. The derivation could continue up to the third derivative, but to avoid overly cumbersome expressions, this work proceeds with a third order approximation by numerically differentiating the analytical second derivatives.
	
	Numerically, one has to decide when to switch from the direct equation \eqref{eq:p_compound_2stream} to the approximation \eqref{eq:p_taylor}. This can be achieved by imposing a tolerance on $\beta$ (or $\Ma_{eq}$), e.g., using the approximation when $-\epsilon < \beta < \epsilon$, with $\epsilon$ an arbitrary small positive number (cf. \cite{restrepo2022viscous}). An alternative condition is based on an estimate of the truncation error in the Taylor series, which can be estimated as the magnitude of the first term after the truncation. The resulting truncation errors on the numerator and the denominator, denoted by $\delta N$ and $\delta \beta$, can be propagated to the pressure gradient as follows:
	\begin{equation}
		\delta \left(\frac{1}{p}\frac{dp}{dx}\right) = \sqrt{\left(\frac{1}{\beta}\right)^2\left(\delta N\right)^2 + \left(-\frac{N}{\beta^2}\right)^2\left(\delta \beta\right)^2}\,. \label{eq:uncert}
	\end{equation}
	Equation \eqref{eq:uncert} is analytically available for the first order approximation using the second derivatives for $\delta N$ and $\delta \beta$ (cf. appendix \ref{sec:dp_dx_second_order}). The second order approximation requires third derivatives to estimate its error, which can be computed with finite differences (as in the quasi-third order approximation). The switch from the direct equation \eqref{eq:p_compound_2stream} to the approximation \eqref{eq:p_taylor} is based on the relative error on the static pressure gradient:
	\begin{equation} \label{eq:dp_dx_blended}
		\frac{1}{p}\frac{dp}{dx} = 
		\begin{cases}
			\text{equation \eqref{eq:p_taylor}} & \text{if } \dfrac{\delta \left(\frac{1}{p}\frac{dp}{dx}\right)}{\left(\frac{1}{p}\frac{dp}{dx}\right)} \leq 5 \% \,, \\
			\text{equation \eqref{eq:p_compound_2stream}} & \text{otherwise.}
		\end{cases}
	\end{equation}
	
	Examples of the distribution of the pressure gradient using the approximations of increasing order are shown in figure \ref{fig:dpdx}. The higher order approximations are more accurate in a wider region around the sonic section. The estimated truncation error for the quasi-third order approximation is not shown. The compound equations have been integrated using the blended equation \eqref{eq:dp_dx_blended} for the pressure. More details on the numerical integration are provided in Section \ref{sec:numerical_solution}.
	
	The blended function \eqref{eq:dp_dx_blended} matches the direct equation \eqref{eq:p_compound_2stream} at the inlet and switches to the approximation \eqref{eq:p_taylor} as the truncation error, computed with \eqref{eq:uncert}, drops below 5 \%. The flow then passes through the sonic section using the approximation \eqref{eq:p_taylor}, effectively avoiding the indeterminate form \eqref{eq:p_delhopital} of the direct equation. Downstream, the direct equation is integrated when the threshold on the truncation error is exceeded again.
	
	In this study, the quasi-third order approximation is used with a backward second order finite difference formula, with a threshold of 5 \% on the second order approximation, since the fourth derivatives needed for the truncation error are not available.
	
	\section{Numerical solution} \label{sec:numerical_solution}
	The governing equations \eqref{eq:pt_i}, \eqref{eq:Tt_i}, \eqref{eq:A_i}, \eqref{eq:p_compound_2stream}, with the approximation \eqref{eq:p_taylor} near the sonic point, form a system of ODEs in function of the axial coordinate $x$, which can thus be integrated in space. The fourth order Runge-Kutta method is used in this study, which is readily available in Python with the `solve\_ivp' function in SciPy\footnote{https://docs.scipy.org/doc/scipy/reference/generated/scipy.integrate.solve\_ivp.html}. In this study, the mesh consists of 100 equally distributed points, which has been found sufficient to accurately respect the conservation equations.
	
	The boundary conditions consist of the total pressures $p_{t,p}, p_{t,s}$, total temperatures $T_{t,p}, T_{t,s}$ and the cross-sections $A_{p}, A_{s}$ at the inlet ($x=0$) and of the static pressure $p_b$ at the outlet ($x=L$), as indicated in figure \ref{fig:sketch}. The distribution of the channel's cross-section $A(x)$ is prescribed. Integrating the system of ODEs requires an `initial' condition at the inlet or at any point from which the integration proceeds forward in $x$. In a single stream computation, the sonic condition $N^*=0$ is used to identify the sonic section and split the integration for the Mach number from the inlet to the sonic section and from the sonic section to the outlet (as proposed by \cite{beans1970computer} and \cite{restrepo2022viscous}).
	
	This is not a viable option for compound flows since the sonic state is not uniquely defined (as $\Ma=1$), but can consist of various combinations of cross-sections $A_i$ and Mach numbers $\Ma_i$. The system state is available at the inlet except for the static pressure $p(0)$. This was found with a shooting method until the boundary condition at the outlet was respected. Firstly, bounds for the inlet pressure are established, considering that $p(0)$ can not exceed the total pressure of any stream by definition, that is the upper bound of $p(0)$ is
	
	\begin{equation}
		p_{max} = \min\left(p_{t,p}(0), p_{t,s}(0)\right)\,.
	\end{equation}
	
\begin{figure}
		\centerline{\includegraphics[width=\linewidth]{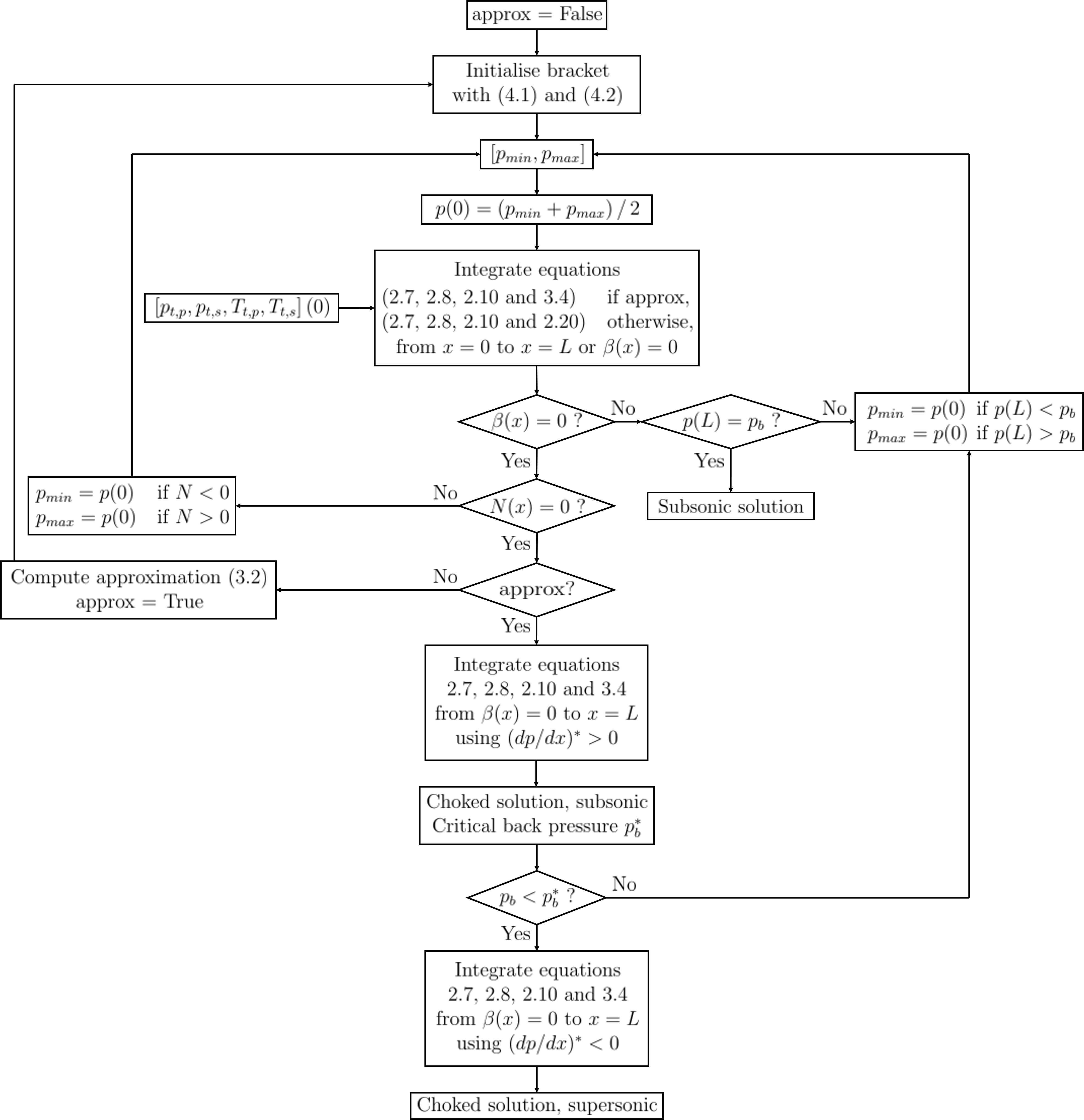}}
		\caption{Flowchart of the numerical procedure to solve the compound nozzle flow. The static pressure $p(0)$ at the inlet is found with a bisection method. A subsonic solution is accepted if the computed back pressure $p(L)$ matches the boundary condition $p_b$. A choked flow requires both $\beta$ and the numerator $N$ to equal zero (see \eqref{eq:p_delhopital}). The approximation \eqref{eq:p_taylor} is computed after convergence of the choked procedure, which is then repeated, replacing \eqref{eq:p_compound_2stream} by \eqref{eq:dp_dx_blended}.}
		\label{fig:flowchart_numerical_solution}
	\end{figure}

	The compound flow is assumed to be subsonic at the inlet (as in a classic nozzle flow), so the second limiting condition is a sonic flow at the inlet ($\beta(0) = 0)$. Using \eqref{eq:beta} and \eqref{eq:Mach_f_pt_p}, the sonic condition at the inlet leads to the lower bound $p_{min}$ for the static pressure at the inlet:
	\begin{equation}
		\beta(0) = 0 = \sum_{i\in[p,s]} \dfrac{A_{i}}{\gamma}\left(\left[\frac{2}{\gamma-1}\left(\left(\frac{p_{t,i}(0)}{p_{min}}\right)^\frac{\gamma-1}{\gamma}-1\right)\right]^{-1} - 1\right)\,,
	\end{equation}
	where $p_{min}$ can be found iteratively. The bounds $\left[p_{min}, p_{max}\right]$ allow to find the correct inlet pressure $p(0)$ with a bisection method, which iteratively restricts the bounds of the interval as depicted in figure \ref{fig:flowchart_numerical_solution}. In each iteration, the governing equations are integrated using the mean of the current interval, after which either its bottom or top half is used in the next iteration. This update depends on whether the computed flow is choked.
	
	In the subsonic case, the inlet pressure $p(0)$ and the computed outlet pressure $p(L)$ are positively correlated: a lower inlet pressure leads to a lower outlet pressure. If the computed back pressure is too low with respect to the boundary condition ($p(L) < p_b$), the inlet pressure should increase and the average pressure $p(0)$ becomes the new lower bound of the bracket in the next iteration. The opposite is true if $p(L) < p_b$. If the pressures $p(L)$ and $p_b$ match, the solution has been found and the flow is compound-subsonic.
	
	If the computed flow becomes sonic ($\beta(x) = 0$), the integration is stopped locally at the sonic point and the numerator $N$ is computed with \eqref{eq:numerator}. If the condition $N=0$ is not respected (see \eqref{eq:p_delhopital}), the inlet pressure $p(0)$ must be updated. A negative numerator indicates that a subsonic flow tends to expand. If this occurs in the sonic section, the sonic point is reached too far upstream, and hence the inlet pressure should be increased. Considering, for example, an isentropic compound flow ($f_{ps}=f_w=0$), where the numerator $N$ is reduced to the gradient $dA/dx$, it is known that the flow becomes sonic at the throat ($dA/dx = 0$). The case $N= dA/dx < 0$ at the sonic section implies that the sonic point is reached in a convergent section; hence, the inlet pressure should be increased to avoid having a non-physical mass flow rate exceeding the choked one. The same argument holds for a non-isentropic flow.
	
	If $N=0$, the flow is sonic in the correct section, and the approximation \eqref{eq:dp_dx_blended} can be computed with the desired order of accuracy. This slightly modifies the governing equations, so the choked solution also changes. Therefore, the procedure to find the inlet pressure $p(0)$ is repeated, replacing \eqref{eq:p_compound_2stream} with \eqref{eq:dp_dx_blended}.
	
	The resulting choked solution is available from the inlet to the sonic section ($\beta(x) = 0$). The governing equations are then integrated up to the outlet, using the positive root $(dp/dx)^*$ of \eqref{eq:p_quadratic} to obtain the subsonic solution. The complete solution is thus subsonic throughout, except at the sonic point. This limiting case corresponds to the highest back pressure which chokes the flow, which is commonly referred to as the critical back pressure $p_b^*$ in ejector modelling. The final step checks whether the choked solution respects the boundary condition $p_b$. The flow remains choked if the imposed back pressure $p_b$ is lower than the critical value $p_b^*$. Otherwise, the static pressure at the inlet should increase, leading to a subsonic solution. 
	
	The supersonic solution can be obtained with the negative root $(dp/dx)^*$ of \eqref{eq:p_quadratic}. If the back pressure $p_b$ lies between the pressures corresponding to the two continuous choked solutions, a shock appears in the diffuser as in the case of a nozzle flow with a single stream. The shock position can be found iteratively by imposing the computed back pressure $p(L)$ to match $p_b$ (see, for example, \cite{restrepo2022viscous}). However, to the authors' knowledge, a compound shock model has not been presented in the literature.
	
	\section{Validation test cases with axisymmetric simulations} \label{sec:numerics}
	The predictions of the 1D theoretical model are compared to post-processed axisymmetric RANS simulations of a compound nozzle in Section \ref{sec:results}. The numerical set-up of these 2D simulations is detailed in Section \ref{sec:numerics_setup}, and the test cases are described in Section \ref{sec:numerics_cases}.
	
	\subsection{Numerical set-up for the RANS validation} \label{sec:numerics_setup}
\begin{figure}
		\centerline{\includegraphics[width=\linewidth]{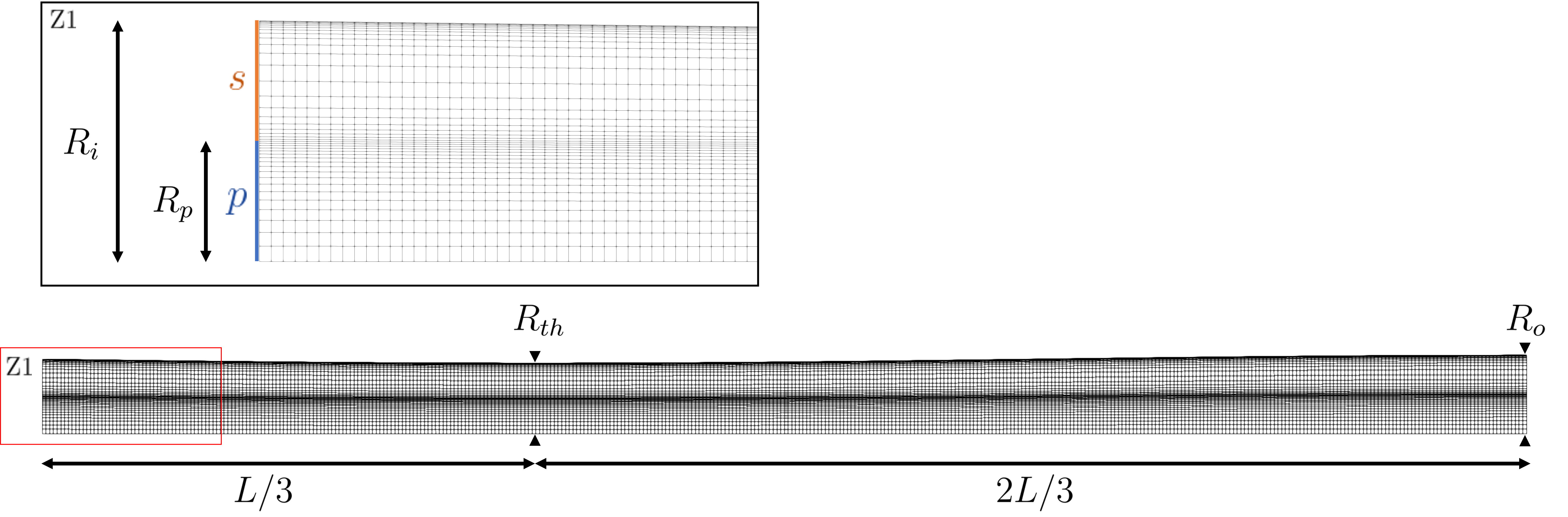}}
		\caption{The computational domain consists of an axisymmetric convergent-divergent nozzle with two distinct inlets. The primary stream (index $p$) is surrounded by the secondary stream (index $s$). The reference geometry is defined by $L = $ 187.5 mm, $R_i = $ 9.5 mm, $R_{th} = $ 9.0 mm, $R_o = $ 10 mm and $R_p = 0.5 R_i$. The model validation considered various throat $R_{th}$ and the primary inlet $R_p$ radii, as described in Section \ref{sec:numerics_cases}, while keeping the same resolution at the wall. The coarse mesh is displayed for visibility.}
		\label{fig:mesh}
	\end{figure}

	The computational domain is shown in figure \ref{fig:mesh}. It consists of an axisymmetric convergent-divergent nozzle with two distinct inlets. The simulation was carried out with a wedge-shaped domain of 5 degrees and a single cell in the tangential direction. The contour of the upper wall was defined as a cosine between $\pi/2$ and $2\pi$ with appropriate scaling to reach the dimensions shown in figure \ref{fig:mesh}:
	\begin{equation} \label{eq:def_geometry}
		r(x) = \dfrac{R_o + R_{th}}{2} + \dfrac{R_o - R_{th}}{2} \cos\left(\dfrac{3\pi}{2} \dfrac{x}{L} + \dfrac{\pi}{2}\right)
	\end{equation}
	
	Hence, the radius at the inlet ($x=0$) follows from the outlet radius $R_o$, the throat radius $R_{th}$ and the length $L$. The throat is located at one-third of the domain. The analysis of the sonic section displacement was focused on a test case with nozzle lengths of 187.5 mm, outlet radius $R_o=$10 mm, inlet radius $R_i = $ 9.5 mm, throat radios $R_{th} = $ 9 mm and $R_p = 0.5 R_i$. For the model validation, several other geometries were considered as reported in more details in Section \ref{sec:numerics_cases}.
	
	The mesh was generated using Gmsh\footnote{https://gmsh.info/} coupled to Python to allow full control of the number of elements and of the refinement near the walls. The simulations were performed with an adapted version of the compressible transient solver rhoCentralFoam in OpenFOAM v9\footnote{https://openfoam.org/version/9/} with the $k-\omega$ SST turbulence model of \cite{menter1993zonal}. The adapted solver does not solve for the tangential component of the velocity, which remains equal to zero (no swirl). \cite{sutherland1893lii}'s law is used to compute the viscosity. The simulations are wall-resolved as shown in Section \ref{sec:numerics_cases}.
	
	The boundary conditions at the inlets consist of the total pressure and total temperature. The primary inlet is chosen to be supersonic, so static pressure is also imposed. A wave transmissive boundary condition is imposed at the exit as the flow is supersonic. The wall is adiabatic and treated with a slip or no-slip condition, depending on the test case. The `wedge' boundary condition is imposed on the sides of the wedge to inform OpenFOAM of the axisymmetry. 
	
	The resulting 2D flow fields are post-processed by identifying the dividing streamline between the streams and by averaging over the resulting cross-sections. The radial position $r_{d}$ of the dividing streamline is found at any point $x$ by integrating the axial mass flux:
	
	\begin{equation} \label{eq:div_streamline}
		\dot{m}_p(x) = \int_0^{r_{d}(x)} \rho(x,r)  u(x,r) \, 2 \pi r \, dr\,,  \,\, \mbox{and } \,\, \dot{m}_s(x) = \int_{r_{d}(x)}^{R} \rho(x,r)  u(x,r) \, 2 \pi r \, dr\,,
	\end{equation}
	where $u$ denotes the axial component of the velocity vector. Knowing the radius $r_d(0)$ of the dividing streamline at the primary inlet, the primary mass flow rate is calculated. The evolution of the dividing streamline at any position $x$ follows from \eqref{eq:div_streamline}. The second step consists of averaging the relevant flow variables across the ejector's section, i.e.
	\begin{align}
		\label{eq:postproc_start}
		\hat{\rho}_{i} (x) &= \frac{1}{A_{i}}\int_{A_{i}}\rho \, dA_{i}\,, \\
		\hat{u}_{i} (x) &= \frac{1}{\hat{\rho}_{i} A_{i}}\int_{A_{i}}\rho u \, dA_{i}\,,\\
		\hat{e}_{t,i} (x) &= \frac{1}{\hat{\rho}_{i} A_{i}}\int_{A_{i}}\rho e_t \, dA_{i}\,,\label{eq:postproc}
	\end{align}
	for $i \in [p,s]$ and where $e_t = c_v T + u^2/2$ denotes the total specific internal energy and $c_v$ is the specific heat at constant volume. The density-averaged velocity and internal energy ensure that the 1D mass flow rate and internal energy match their integrated 2D counterpart. The averaged static temperature, static pressure and Mach number follow from \eqref{eq:postproc_start}-\eqref{eq:postproc} and the specific gas constant $R$:
	\begin{equation}
		\hat{T}_{i} = \frac{1}{c_v} \left(\hat{e}_{t,i}-\frac{1}{2}\hat{u}_{i}^2\right)\,,  \quad \quad \hat{p}_i = \hat{\rho}_i R \hat{T}_i\,,  \quad \mbox{and } \quad
		\hat{\Ma}_{i} = \frac{\hat{u}_{i}}{\sqrt{\gamma R \hat{T}_{i}}}\,.\label{eq:1D_Mach}
	\end{equation}
	
	\subsection{Selected Test Cases} \label{sec:numerics_cases}
	Three simulations were carried out where the wall and inter-stream friction forces were numerically (de)-activated. The boundary conditions are listed in table \ref{tab:boundary_conditions}. At both inlets, the turbulent mixing length was set equal to 7 \% of the hydraulic diameter, as recommended by \cite{versteeg2007introduction}. The turbulence intensity was set to 1 \%, as this value proved to best match the friction coefficients predicted by \eqref{eq:f_ps_correlation}.
	
\begin{figure}
		\center
		\begin{subfigure}[t]{0.48\linewidth}
			\includegraphics[width=\linewidth]{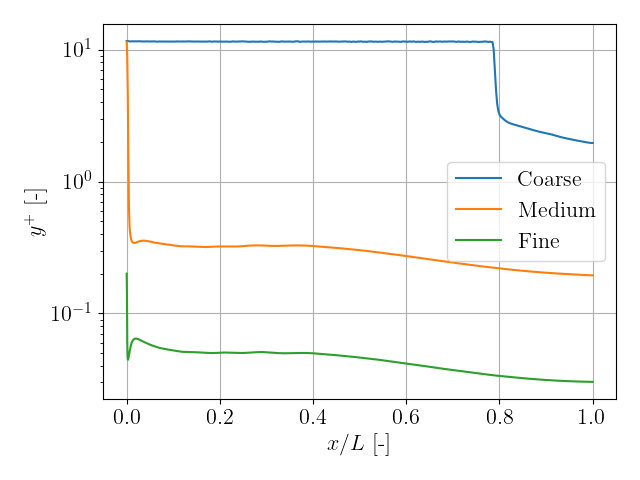}
			\subcaption{$y^+$ of the first cell near the wall.}
			\label{fig:mesh_conv_yplus}
		\end{subfigure}
		\begin{subfigure}[t]{0.48\linewidth}
			\includegraphics[width=\linewidth]{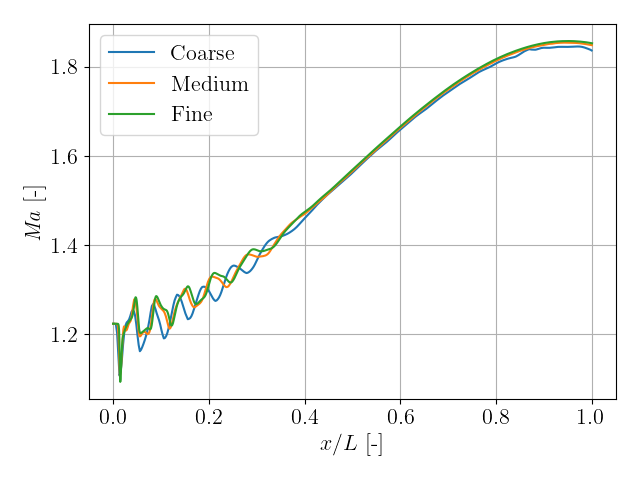}
			\subcaption{Mach number at the centerline.}
			\label{fig:mesh_conv_Mach}
		\end{subfigure}
		\caption{The coarse, medium and fine mesh contain $400 \times 25$, $800 \times 52$ and $1600 \times 106$ elements respectively. The medium and fine meshes lead to a wall-resolved simulation and to good agreement for the Mach number at the centerline. The fine mesh is retained for the other simulations and the post-processing.}
		\label{fig:mesh_conv}
	\end{figure}

	The first simulation is inviscid and thus serves as a reference for the original compound flow theory without momentum exchange. The second simulation is viscous with a slip condition at the wall: only friction forces between the streams are thus active in this case. Finally, wall friction is included in the third case by imposing a no-slip condition along the upper boundary. A mesh convergence study was carried out on the viscous simulation with the no-slip condition. The number of cells was doubled in each direction, leading to meshes with $400 \times 25$, $800 \times 52$ and $1600 \times 106$ cells. The resulting distributions of $y^+$ near the wall and the Mach number at the centerline are shown for the three meshes in figure \ref{fig:mesh_conv}. The medium and fine meshes show good agreement and are both wall-resolved. The wall friction Reynolds number $\Rey_{\tau} = \rho u_{\tau} D_h / \mu$ equals $1.9\cdot 10^4$ at the inlet and $5.3 \cdot 10^3$ at the outlet in case 3. The results on the fine mesh are used in the next section to profit from its higher resolution.
	
	\begin{table}
		\centering
		\begin{tabular}{cccccccccc}
			Case & $p_{t,p}$ [bar] & $p_p$ [bar] & $T_{t,p}$ [K] & $p_{t,s}$ [bar] & $T_{t,s}$ [K] & $TI$ [\%] & $l_t$ [m] & Wall & Comments\\[3pt]
			$1$ & $3.0$ & $1.21$ & $300$ & $1.5$ & $300$ & n/a & n/a & slip & inviscid \\
			$2$ & $3.0$ & $1.21$ & $300$ & $1.5$ & $300$ & $1$ & $6.65 \cdot 10^{-4}$ & slip & viscous \\
			$3$ & $3.0$ & $1.21$ & $300$ & $1.5$ & $300$ & $1$ & $6.65 \cdot 10^{-4}$ & no-slip & viscous
		\end{tabular}
		\caption{Boundary conditions of the simulations presented in Section \ref{sec:results}. The primary inlet is supersonic with $\Ma_p=1.22$. They respectively cover (1) an isentropic case, (2) a case with friction between the streams but without wall friction and (3) a case with both forces.}
		\label{tab:boundary_conditions}
	\end{table}
	
	A parametric study was carried out to validate the model in a wider range of conditions. The reference case is the viscous simulation with the no-slip condition at the wall (case 3 in table \ref{tab:boundary_conditions}), as it is the most representative for practical applications. Three parameters were varied independently: (1) the total pressure at the secondary inlet between $p_{t,s} =$ 0.5 and 3 bar, (2) the radius of the throat between $R_{th}/R_o = $ 0.6 and 0.9 and (3) the radius of the primary inlet between $R_p/R_i = $ 0.3 and 0.7 (see figure \ref{fig:mesh}). The correlations \eqref{eq:f_w_correlation} and \eqref{eq:f_ps_correlation} were used in all cases, so the model was not tuned further.
	
	The resulting distributions of the static pressure and the Mach numbers depend on the geometry and the inlet conditions, so the static pressure imposed at the supersonic primary inlet changes too. However, the static pressure at the inlet corresponding to a choked flow without a shock train is initially unknown. The 1D model proved to be a useful tool to compute this pressure, as it provides a cross-stream averaged description of a choked flow with a \emph{uniform} static pressure profile. The 1D model was thus run \emph{first}, to provide the boundary conditions for the RANS simulations \emph{afterwards} (see table \ref{tab:var_boundary_conditions}). The primary stream is supersonic in most cases, except if $p_{t,s} \geq 2$ bar. In that case, only the total pressure and total temperature were imposed. The other boundary conditions remained identical to those in table \ref{tab:boundary_conditions}.
	
	\begin{table}
		\centering
		\begin{tabular}{ccccc}
			$p_{t,s}$ [bar] & $R_{th}/R_o$ [-] & $R_p/R_i$ [-] & $p_p$ [bar] & Comments\\[3pt]
			\hline
			1.5 & 0.9 & 0.5 & 1.21 & reference case\\
			\hline
			\textbf{0.5} & 0.9 & 0.5 & 0.40 &  \\
			\textbf{1.0} & 0.9 & 0.5 & 0.81 &  \\
			\textbf{2.0} & 0.9 & 0.5 & n/a & subsonic primary inlet \\
			\textbf{2.5} & 0.9 & 0.5 & n/a & subsonic primary inlet \\
			\textbf{3.0} & 0.9 & 0.5 & n/a & subsonic primary inlet \\
			\hline
			1.5 & \textbf{0.6} & 0.5 & 1.45 &  \\
			1.5 & \textbf{0.7} & 0.5 & 1.40 &  \\
			1.5 & \textbf{0.8} & 0.5 & 1.32 &  \\
			\hline
			1.5 & 0.9 & \textbf{0.3} & 1.16 &  \\
			1.5 & 0.9 & \textbf{0.4} & 1.18 &  \\
			1.5 & 0.9 & \textbf{0.6} & 1.25 &  \\
			1.5 & 0.9 & \textbf{0.7} & 1.30 &  \\
		\end{tabular}
		\caption{Boundary conditions for the parametric study. The primary static pressure $p_p$ was computed a priori with the 1D model and served a posteriori as a boundary condition for the RANS simulations. The primary stream is subsonic at the inlet for $p_{t,s} \geq 2$ bar, so the static pressure is not imposed.}
		\label{tab:var_boundary_conditions}
	\end{table}
	
	\FloatBarrier
	
	\section{Results} \label{sec:results}
	
	The two approaches to close the model from Section \ref{sec:closure} are compared in Section \ref{sec:results_closure}, followed by a validation of the model in various operating conditions and geometries in Section \ref{sec:results_validation}. The effect of the friction forces on compound choking is illustrated in Section \ref{sec:results_comparison}. Compound choking is shown to be a plausibly more general flow blockage mechanism than individual streams reaching a unitary Mach number in Section \ref{sec:results_blockage}. This is demonstrated through a choked compound flow with a fully subsonic secondary stream, both with the 1D model and with an axisymmetric RANS simulation.
	
	\subsection{Model closure} \label{sec:results_closure}
\begin{figure}
		\center
		\begin{subfigure}[t]{0.48\linewidth}
			\includegraphics[width=0.99\linewidth]{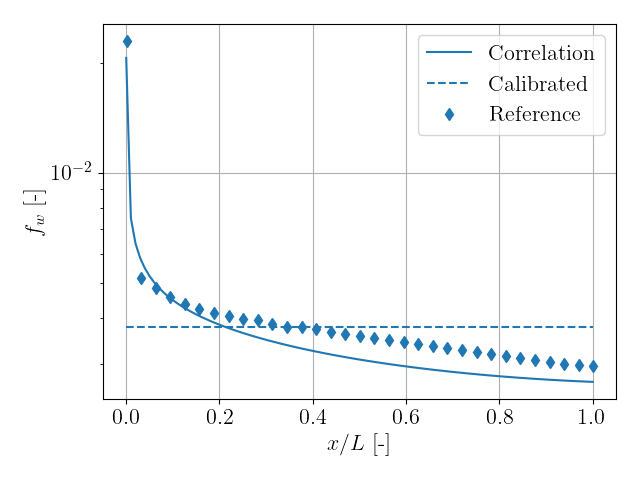}
			\subcaption{Wall friction coefficient $f_w$.}
			\label{fig:comparison_f_w}
		\end{subfigure}
		\begin{subfigure}[t]{0.48\linewidth}
			\includegraphics[width=0.99\linewidth]{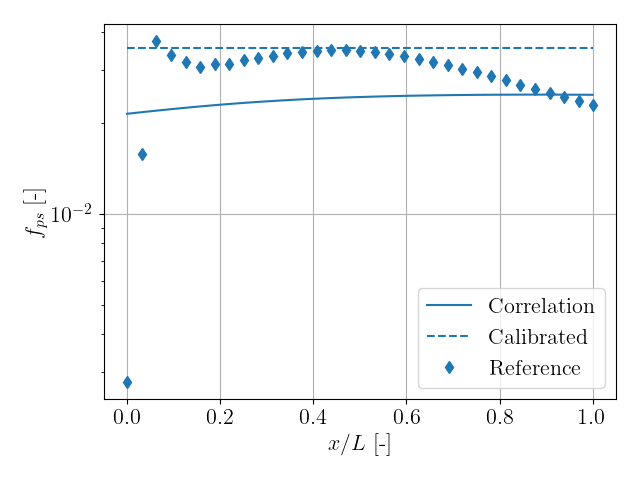}
			\subcaption{Inter-stream friction coefficient $f_{ps}$.}
			\label{fig:comparison_f_ps}
		\end{subfigure}
		\\
		\begin{subfigure}[t]{0.48\linewidth}
			\includegraphics[width=0.99\linewidth]{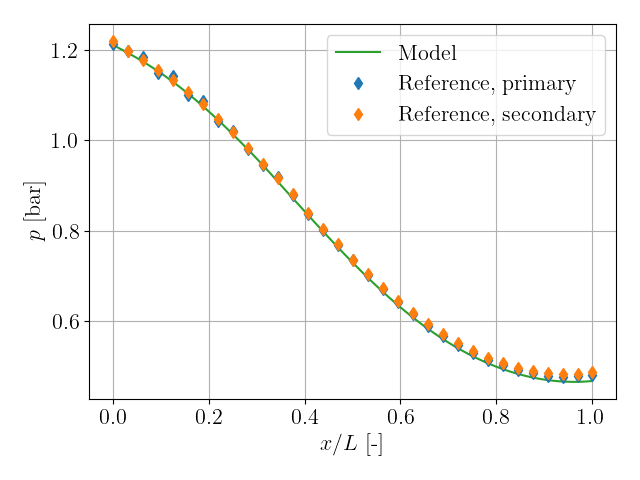}
			\subcaption{Prediction of the static pressure with the correlations.}
			\label{fig:comparison_f_corr}
		\end{subfigure}
		\begin{subfigure}[t]{0.48\linewidth}
			\includegraphics[width=0.99\linewidth]{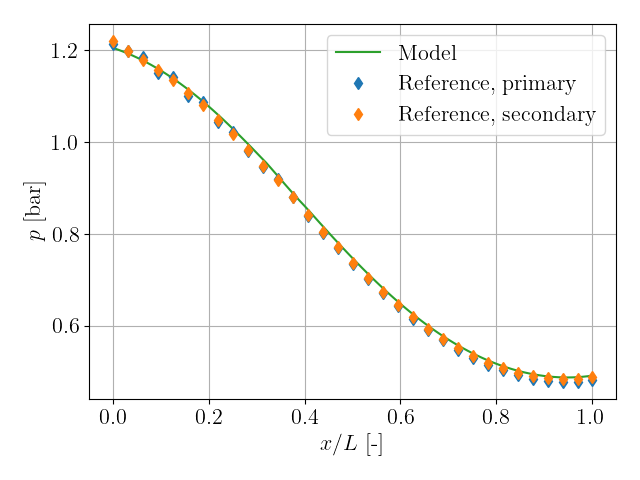}
			\subcaption{Prediction of the static pressure with calibrated constants.}
			\label{fig:comparison_f_calib}
		\end{subfigure}
		\caption{Comparison of the two approaches for determining the friction coefficients (see Section \ref{sec:closure}). The model predicts the static pressure accurately with the correlations \eqref{eq:f_w_correlation} and \eqref{eq:f_ps_correlation}, and with the calibrated constants ($f_w = 0.00377$, $f_{ps}=0.0355$).}
		\label{fig:comparison_f}
	\end{figure}

	Before proceeding with the analysis of the investigated test cases, figure \ref{fig:comparison_f} showcases the impact of the friction coefficients at the wall and at the dividing streamline for the test case 3. Figures \ref{fig:comparison_f_w} and \ref{fig:comparison_f_ps} show the streamwise evolution of the friction coefficients (1) obtained with the correlations \eqref{eq:f_w_correlation} and \eqref{eq:f_ps_correlation}, (2) identified as constants by the inverse method as described in Section \ref{sec:closure} and (3) extracted from the viscous RANS simulations as reference. The correlation \eqref{eq:f_w_correlation} for the wall friction is adequate and follows the correct trend. The calibrated constant matches the reference on average. The friction between the streams is mostly underpredicted by correlation \eqref{eq:f_ps_correlation}, as also observed by \cite{grazzini2015constructal}. This was found to be sensitive to the turbulent intensity at the inlets, which directly impacts the turbulent viscosity and hence the friction between the streams. Analysing this effect in further details is beyond the scope of the current work.
	
	More importantly, the distribution of the static pressure is predicted accurately by the 1D model in both cases, as shown in figures \ref{fig:comparison_f_corr} and \ref{fig:comparison_f_calib}. The averaged static pressure is computed from the RANS simulations with equations \eqref{eq:div_streamline} - \eqref{eq:1D_Mach} and is displayed with diamonds as the reference. For plotting purposes, the spacing in the markers is much larger than the mesh resolution.
	
	It is worth noticing that the static pressure in the two streams oscillates near the inlet due to a weak shock train. This effect is not captured in the 1D model because of the assumption of uniform static pressure. The agreement between the model and the RANS simulations shows that the correlations \eqref{eq:f_w_correlation} and \eqref{eq:f_ps_correlation} are adequate for the purposes of this work. Moreover, the agreement between predictions with both constant and varying coefficients indicates a weak sensitivity to these parameters, justifying the exclusion of their derivatives in the approximation in section \ref{sec:approx}. All the results presented for the 1D model in the remaining of this section were obtained with the correlations. This approach has the advantage of not requiring calibration since the model is fully closed when using \eqref{eq:f_w_correlation} and \eqref{eq:f_ps_correlation}.
	
	\subsection{Model validation}\label{sec:results_validation}		
		
\begin{figure}
		\center
		\begin{subfigure}[t]{0.9\linewidth}
			\includegraphics[width=\linewidth]{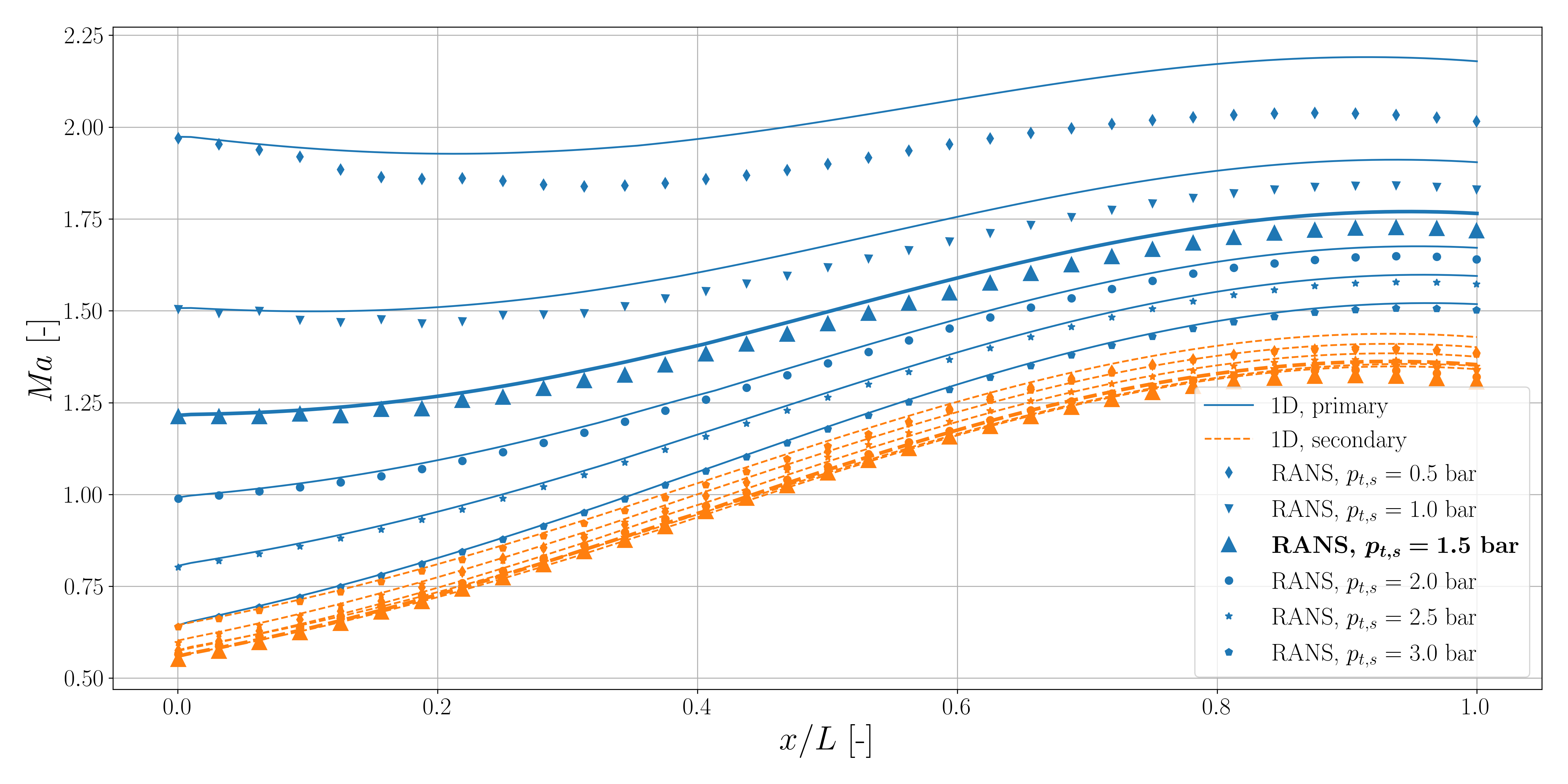}
			\subcaption{Mach numbers.}
			\label{fig:varpts_Mach}
		\end{subfigure}
		\\
		\begin{subfigure}[t]{0.9\linewidth}
			\includegraphics[width=\linewidth]{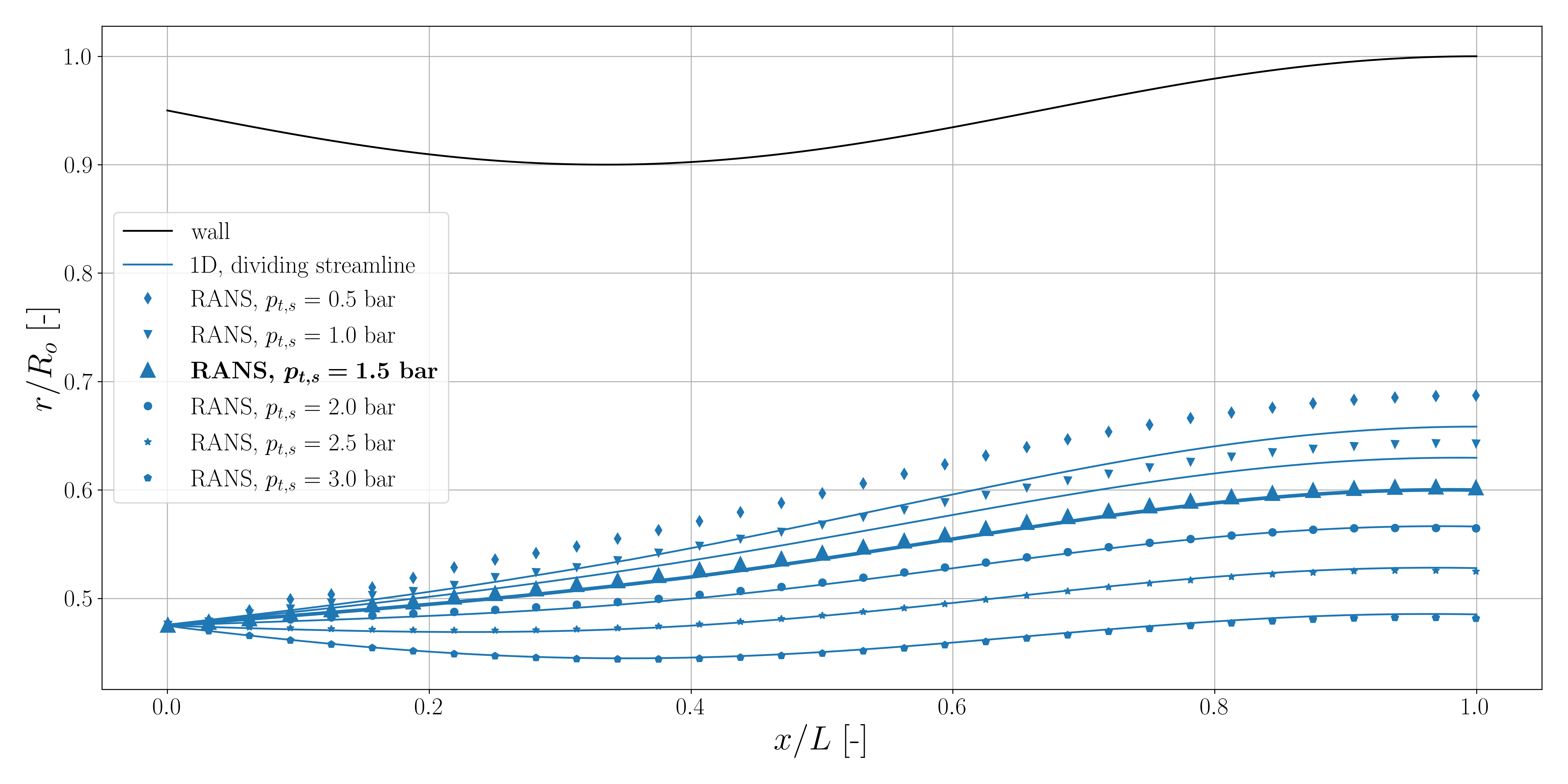}
			\subcaption{Dividing streamlines.}
			\label{fig:varpts_rdiv}
		\end{subfigure}
		\caption{Comparison of the model predictions and the post-processed RANS simulations with varying total pressure $p_{t,s}$ at the secondary inlet and constant total pressure $p_{t,p} = 3$ bar at the primary inlet. The reference case is highlighted in bold. The geometry is identical to the reference case in Section \ref{sec:numerics_setup}. The primary Mach number decreases with increasing $p_{t,s}$ due to a higher static pressure at the inlet and eventually becomes subsonic. The agreement degrades with a larger difference in inlet pressures.}
		\label{fig:varpts}
	\end{figure}

	Figure \ref{fig:varpts} illustrates the distribution of the Mach numbers and the dividing streamlines for the different secondary inlet pressures (see table \ref{tab:var_boundary_conditions}). The primary Mach number decreases significantly with increasing total pressure $p_{t,s}$, due to an increased static pressure at the inlet. Similarly, the secondary Mach number increases with increasing $p_{t,s}$, but is less sensitive. When the total pressures are equal (uniform flow), the Mach numbers are equal in each stream at the inlet. Downstream, the secondary Mach number decreases relative to the primary due to wall friction, which is not fully counteracted by the weak shear force from the quasi-uniform flow.

	The model agreement with RANS simulations worsens as the inlet pressure difference grows. Notably, in these cases, the primary Mach number and dividing streamline oscillate near the inlet, due to the formation of a mild shock train. The resulting oblique shocks introduce pressure losses that are not captured by the 1D model, which assumes a uniform static pressure across the streams. Additionally, the maximal relative error of the model amounts to 10 \%, which is comparable to the scatter of the data on which the correlation \eqref{eq:f_ps_correlation} is based (see \cite{papamoschou1993model}). Despite this, the overall agreement is satisfactory, especially considering that the model was not specifically calibrated for these conditions.
		
\begin{figure}
		\center
		\begin{subfigure}[t]{0.9\linewidth}
			\includegraphics[width=\linewidth]{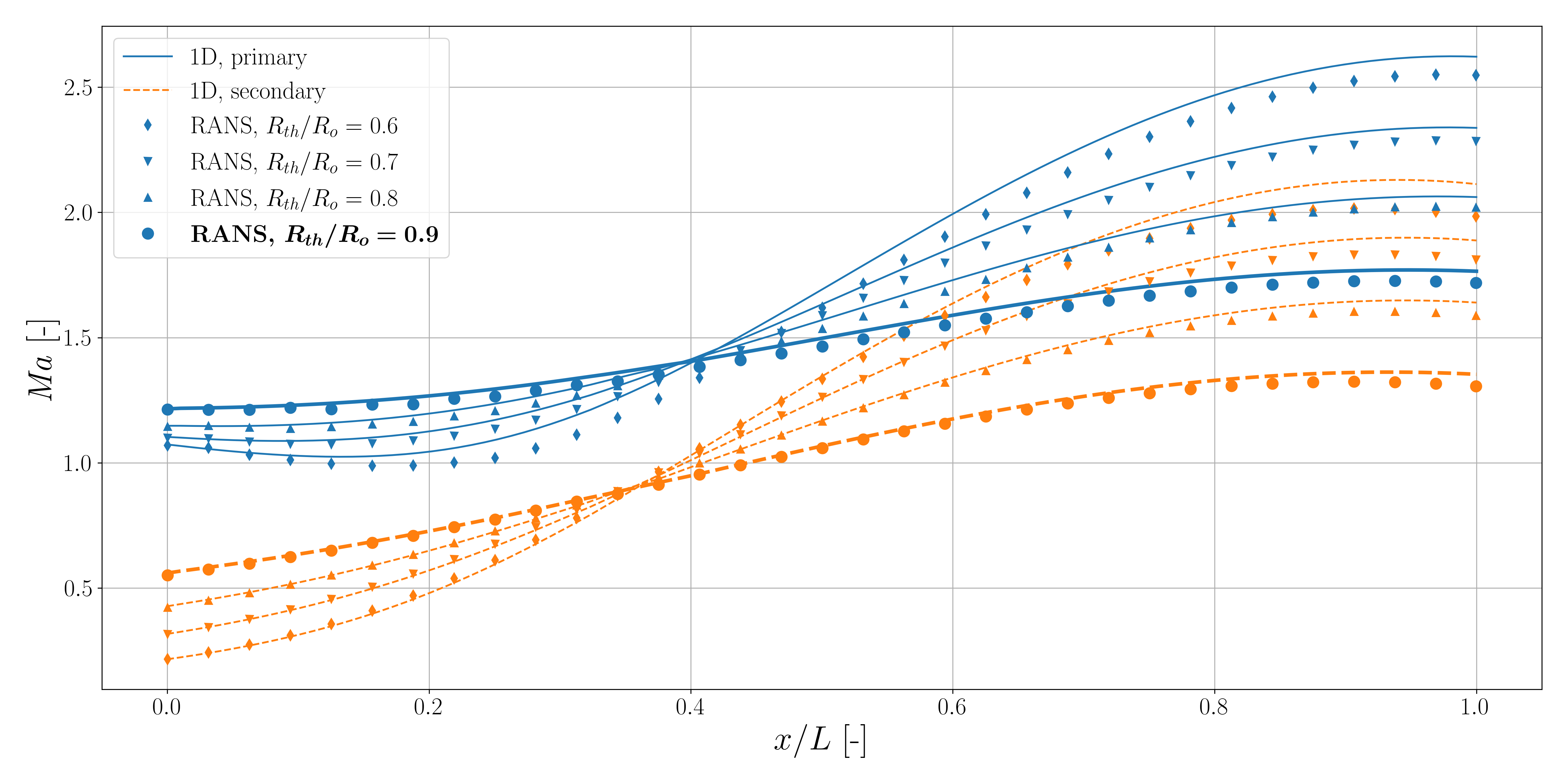}
			\subcaption{Mach numbers.}
			\label{fig:varthroat_Mach}
		\end{subfigure}
		\\
		\begin{subfigure}[t]{0.9\linewidth}
			\includegraphics[width=\linewidth]{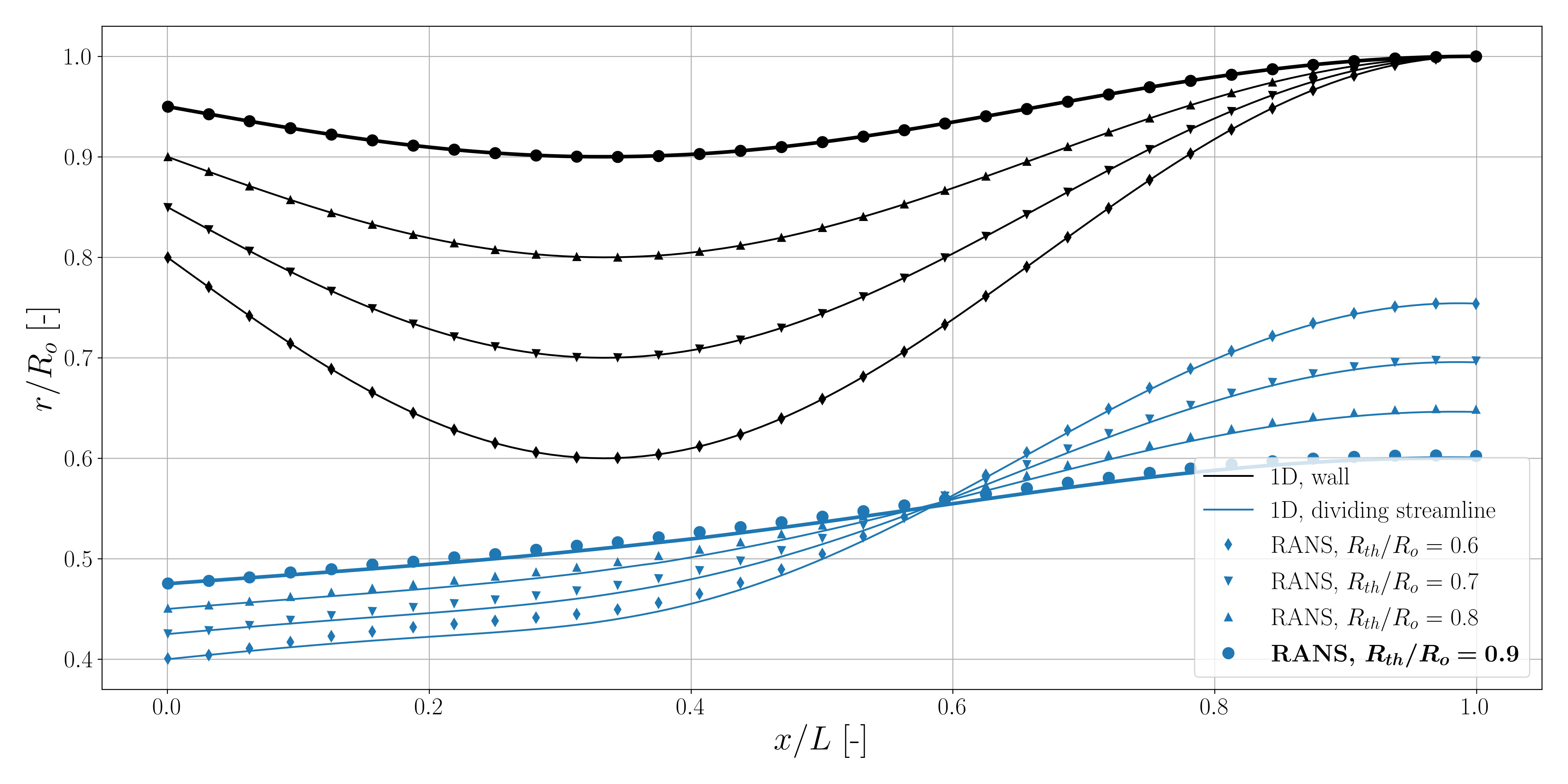}
			\subcaption{Dividing streamlines.}
			\label{fig:varthroat_rdiv}
		\end{subfigure}
		\caption{Comparison of the model predictions and the post-processed RANS simulations with varying radius of the throat. The reference case is highlighted in bold. The total pressures at the inlet are identical to the reference case 3 in table \ref{tab:boundary_conditions}. The Mach numbers decrease at the inlet and increase at the outlet with a narrowing throat. The curvature becomes important for the smallest throat, inducing 2D effects and reducing the accuracy of a 1D formulation.}
		\label{fig:varthroat}
	\end{figure}

	The throat radius $R_{th}$ was varied between 60 \% and 90 \% of the outlet radius $R_o$, while maintaining the same total pressures as in case of table \ref{tab:boundary_conditions}. A narrower throat induces a stronger expansion, resulting in lower Mach numbers upstream and higher Mach numbers downstream of the sonic section (see figure \ref{fig:varthroat}). The agreement between the model and the RANS simulations is excellent, with discrepancies below 10 \%. The overpredicted Mach numbers at the outlet are attributed to underpredicted friction coefficients by \eqref{eq:f_w_correlation} and \eqref{eq:f_ps_correlation} with respect to the RANS simulations, as also observed in figures \ref{fig:comparison_f_w} and \ref{fig:comparison_f_ps}. It is worth noticing that the radius of the wall at the inlet changes with the throat due to the cosine-shaped profile between $\pi/2$ and $2 \pi$, as defined by \eqref{eq:def_geometry}. The primary inlet changes accordingly and extends halfway to the total inlet in each configuration.

\begin{figure}
		\center
		\begin{subfigure}[t]{0.9\linewidth}
			\includegraphics[width=\linewidth]{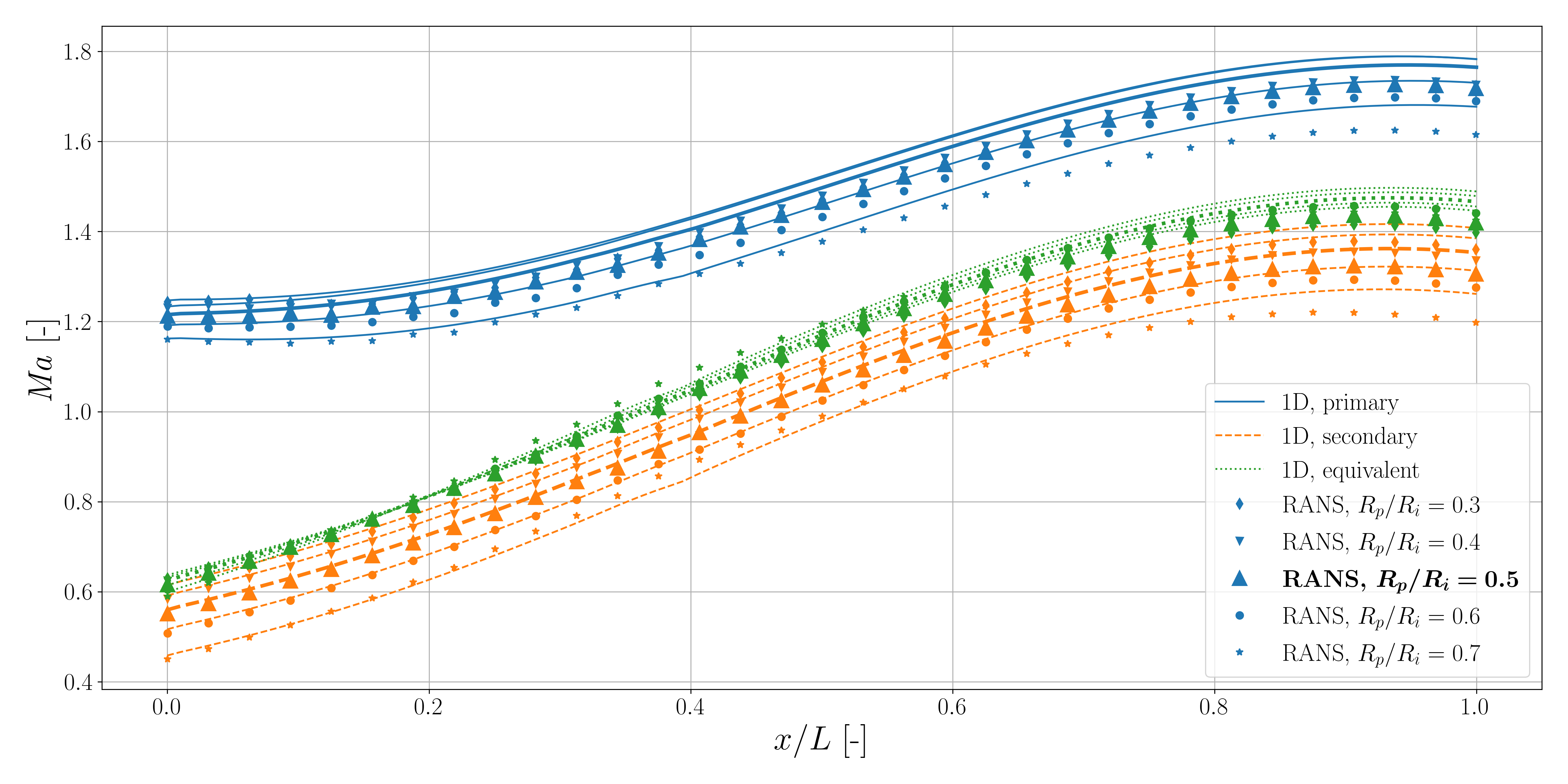}
			\subcaption{Mach numbers.}
			\label{fig:varinlet_Mach}
		\end{subfigure}
		\\
		\begin{subfigure}[t]{0.9\linewidth}
			\includegraphics[width=\linewidth]{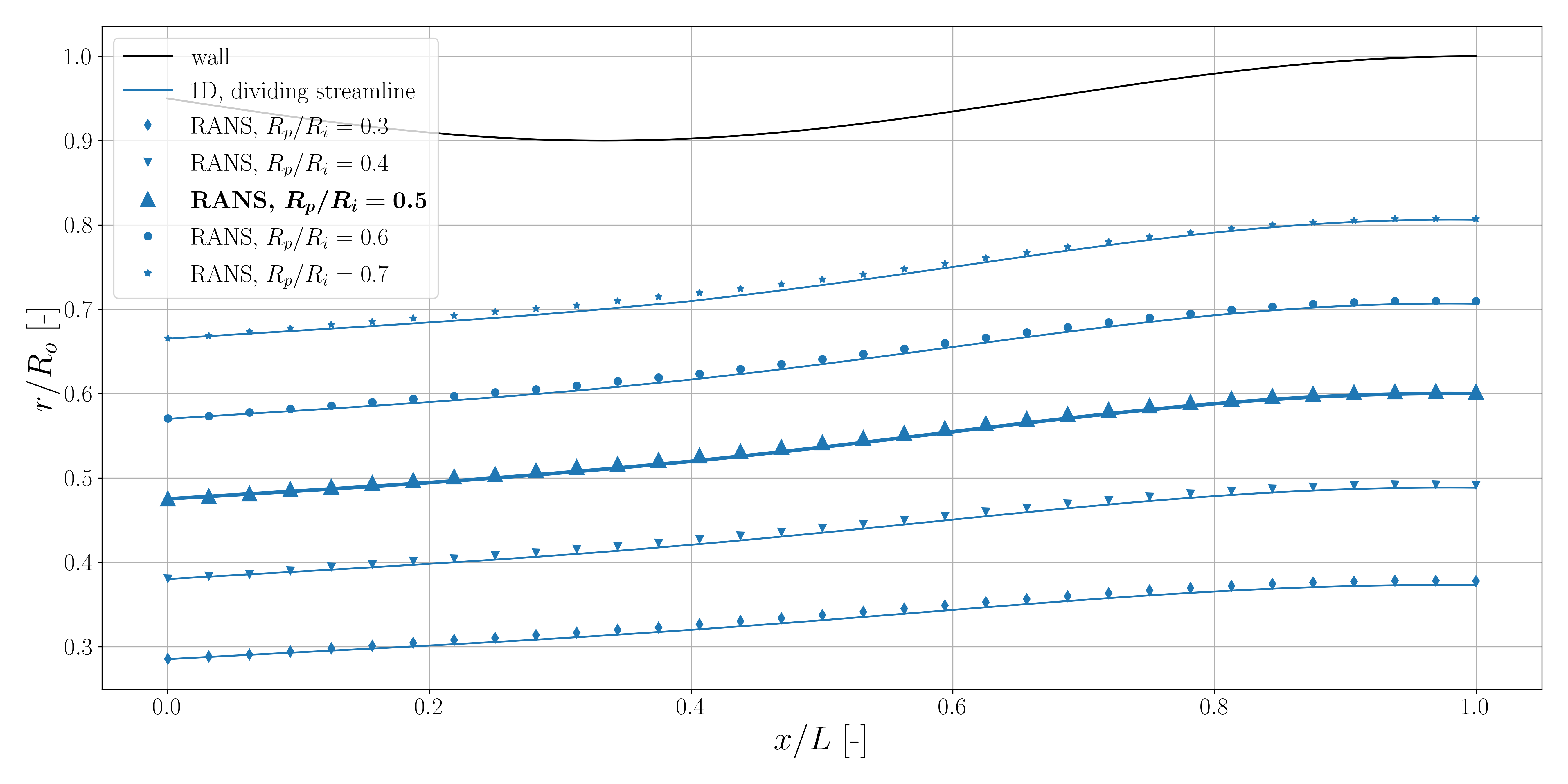}
			\subcaption{Dividing streamlines.}
			\label{fig:varinlet_rdiv}
		\end{subfigure}
		\caption{Comparison of the model predictions and the post-processed RANS simulations with varying radius of the primary inlet. The reference case is highlighted in bold. The total pressures at the inlet are identical to the reference case 3 in table \ref{tab:boundary_conditions}. A larger primary cross-section at the inlet increases the average total pressure at the inlet and hence increases the static pressure required for choking. Consequently, both Mach numbers decrease with an increasing radius of the primary inlet. }
		\label{fig:varinlet}
	\end{figure}

	Finally, the radius of the primary inlet is varied between 30 and 70 \% of the total radius at the inlet, while keeping the same profile of the wall and the same total pressures as the reference case. Figure \ref{fig:varinlet} shows that both Mach numbers decrease with an increasing radius of the primary inlet. This is due to a higher average total pressure at the inlet (a larger portion with $p_{t,p} = 3$ bar instead of $p_{t,s} = 1.5$ bar), which in turn leads to a higher static pressure of the choked flow. Despite the lower Mach numbers in both streams, the equivalent Mach number is comparable. This surprising effect is possible through the non-linear relation between the Mach numbers and the cross-sections in the terms of $\beta$ in \eqref{eq:beta}. Clearly, the Mach numbers alone do not suffice to conclude on the state of the flow and should be complemented with information on the cross-sections. The 1D model captures these trends accurately, although the Mach numbers are overpredicted near the outlet as observed earlier in figure \ref{fig:varthroat}.

	\subsection{The effect of friction on compound choking} \label{sec:results_comparison}
\begin{figure}
		\centerline{\includegraphics[width=\linewidth]{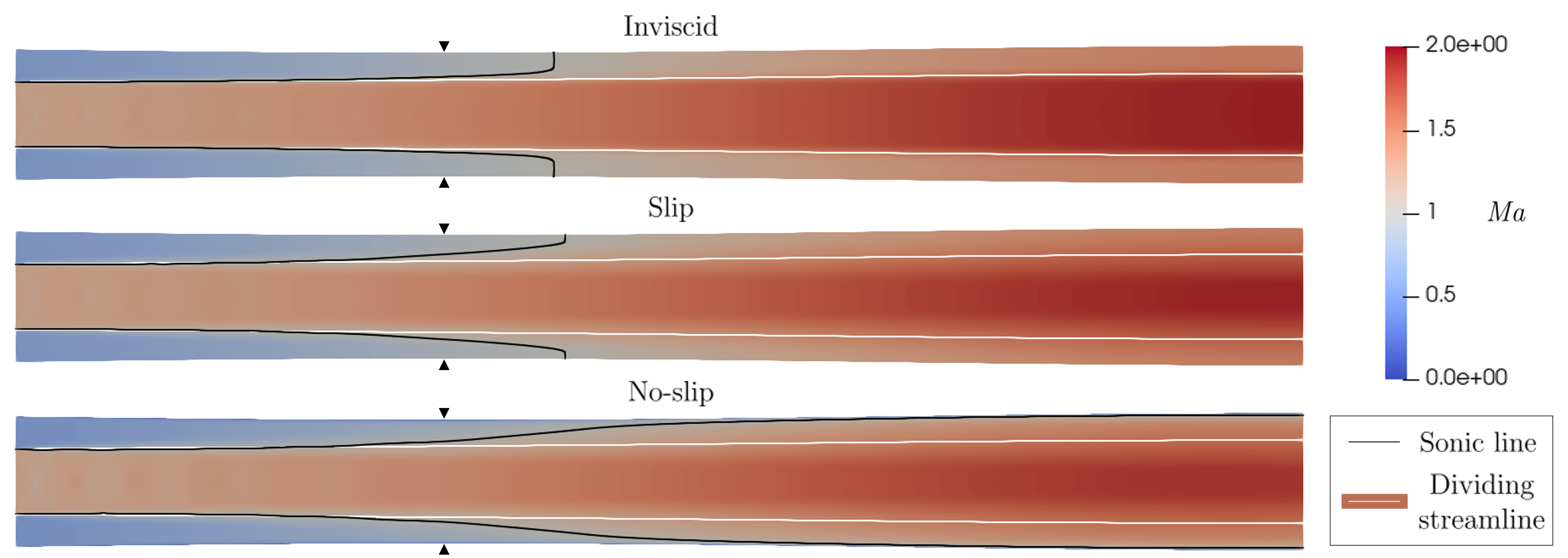}}
		\caption{Mach number contours of the three simulations (cf. table \ref{tab:boundary_conditions}). The black line indicates the sonic line and tends towards the wall as the compound flow becomes supersonic. The dividing streamline is indicated in white. The arrows indicate the geometrical throat. The numerical domain depicted in figure \ref{fig:mesh} has been mirrored for visibility.}
		\label{fig:2D_CFD}
	\end{figure}

	Figure \ref{fig:2D_CFD} shows the Mach number fields of the three simulations in table \ref{tab:boundary_conditions}, along with the dividing streamline in white and the sonic line in black. The shear layer in the inviscid case is only due to numerical diffusion, so the two streams remain clearly distinct across the dividing streamline. The sonic line ends on the wall downstream of the geometrical throat (indicated by the arrows). The primary stream is thus supersonic at the throat and the secondary stream is still subsonic. The secondary stream is accelerated through shear in the viscous simulations, causing the sonic line to move radially outwards. The sonic line no longer reaches the wall when a no-slip condition is imposed. The development of the boundary layer tends to slow down the flow, leading to a lower maximal Mach number.
	
\begin{figure}
		\center
		\begin{subfigure}[t]{0.48\linewidth}
			\includegraphics[width=0.9\linewidth]{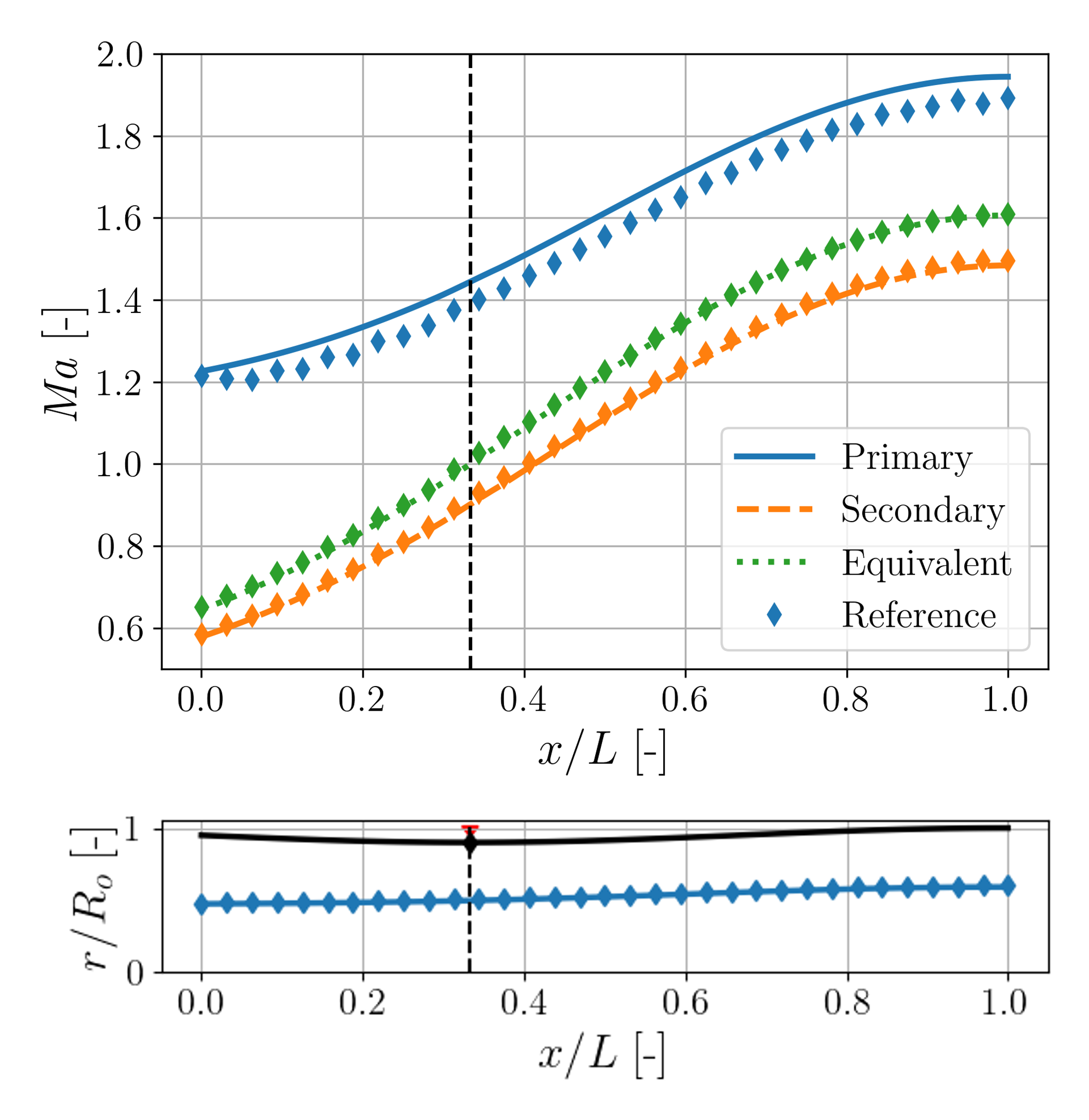}
			\subcaption{Case 1, sonic at $x/L = 0.333$.}
			\label{fig:1D_nozzle_inviscid}
		\end{subfigure}
		\\
		\begin{subfigure}[t]{0.48\linewidth}
			\includegraphics[width=0.99\linewidth]{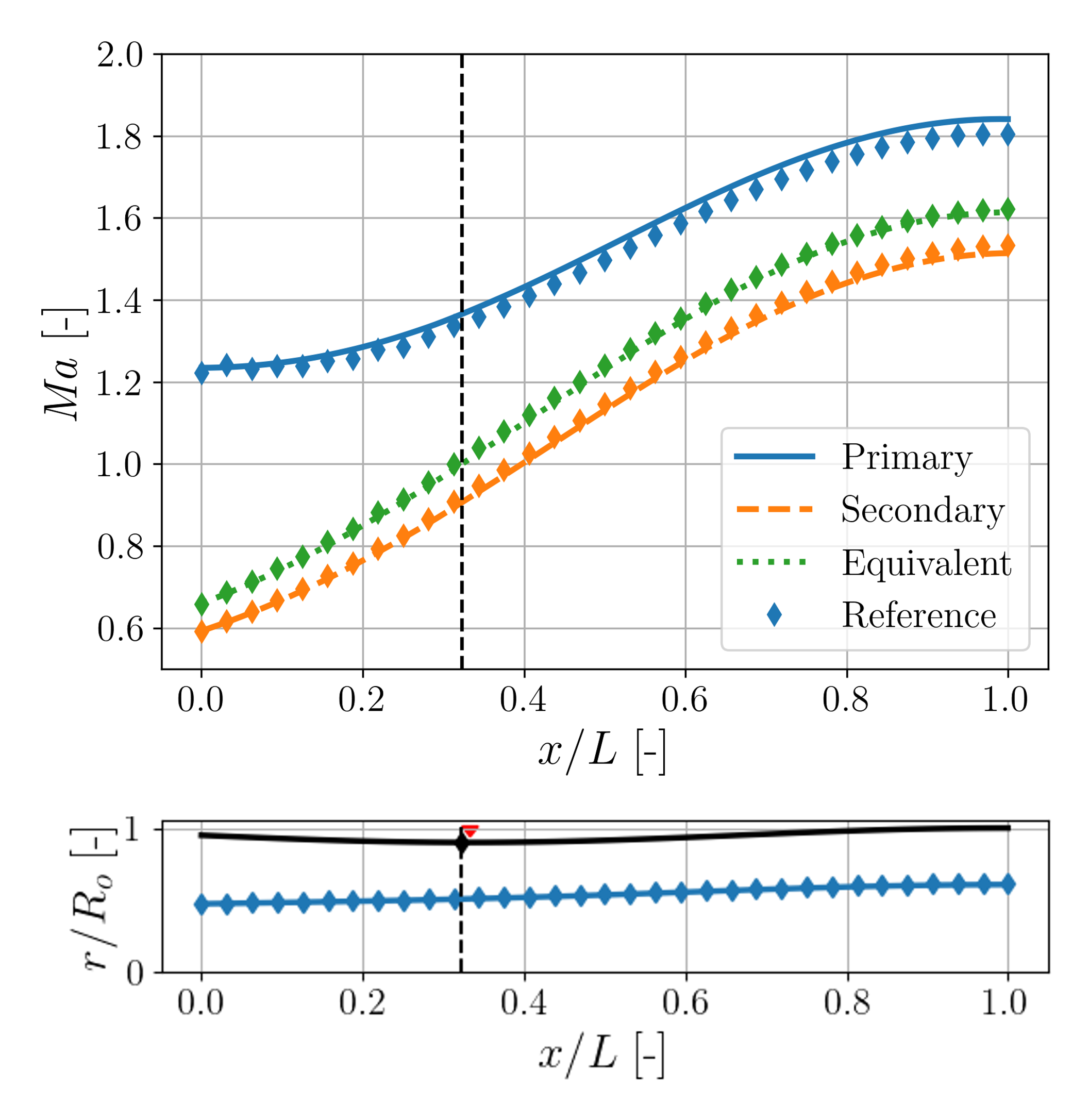}
			\subcaption{Case 2, sonic at $x/L = 0.323$.}
			\label{fig:1D_nozzle_slip}
		\end{subfigure}
		\begin{subfigure}[t]{0.48\linewidth}
			\includegraphics[width=0.99\linewidth]{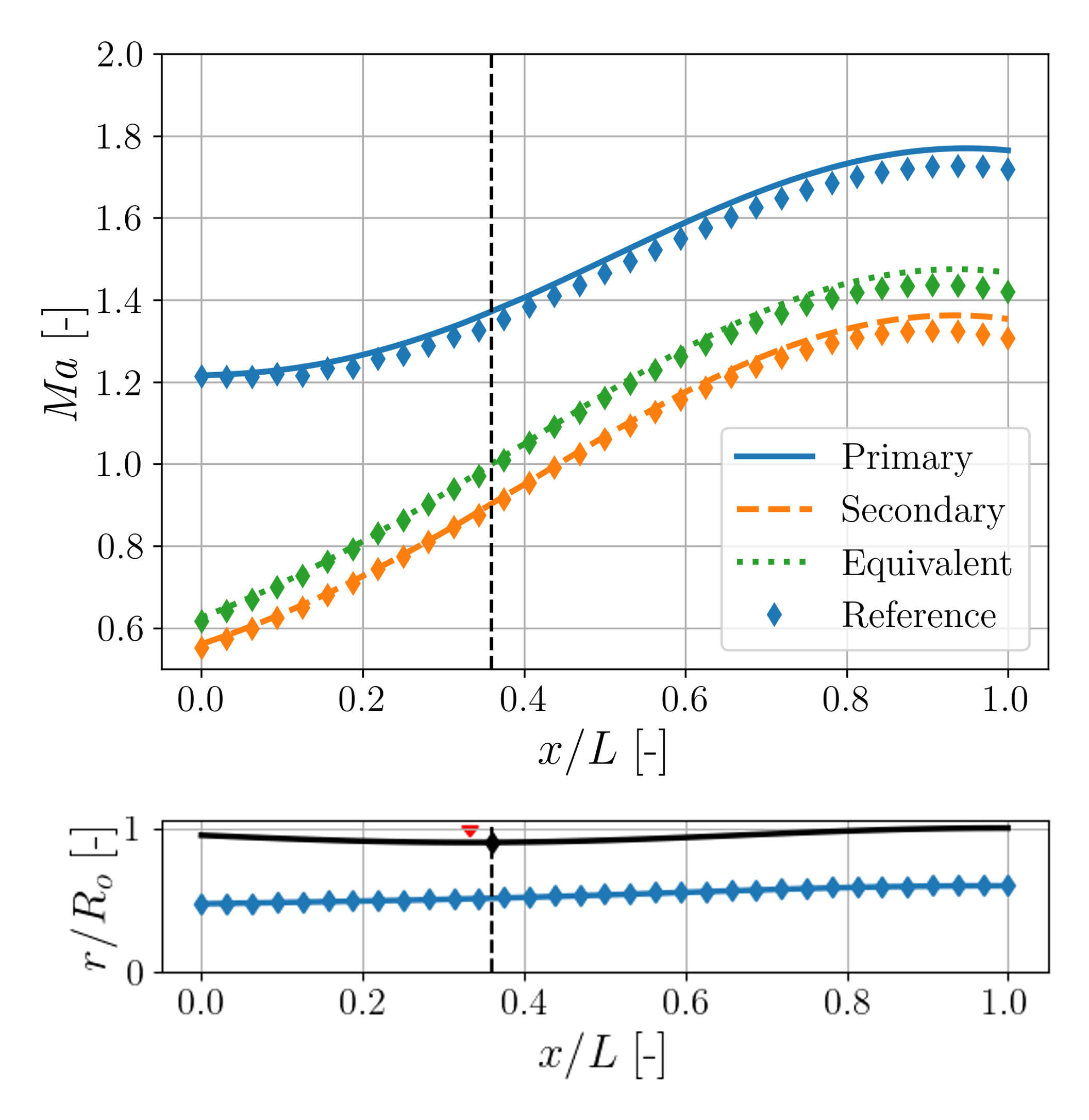}
			\subcaption{Case 3, sonic at $x/L = 0.359$.}
			\label{fig:1D_nozzle_noslip}
		\end{subfigure}
		\caption{Comparison of the Mach number computed with the 1D compound flow theory and from the post-processed RANS results. The sonic section (dotted line) coincides with the throat (red triangle) in the inviscid case 1. Friction between the streams pushes the sonic section upstream (case 2), while wall friction pushes it downstream in case 3 (see \eqref{eq:numerator}). The used friction coefficients are given by \eqref{eq:f_w_correlation} and \eqref{eq:f_ps_correlation}.}
		\label{fig:1D_nozzle}
	\end{figure}

	The averaged Mach numbers extracted from the RANS simulations are displayed with diamonds in figure \ref{fig:1D_nozzle} as the reference. The predictions of the proposed model are shown in full lines. It is worth noticing that the static pressure at the inlet for the 1D model prediction is not imposed to be equal to the one from the RANS simulations, but rather results from the iterative procedure laid out in Section \ref{sec:approx}. The only difference in the three cases is the friction coefficients, which are used to obtain the 1D predictions. In the inviscid case, both naturally equal 0. The friction coefficients are computed with \eqref{eq:f_w_correlation} and \eqref{eq:f_ps_correlation} for the viscous simulations.
	
	The key finding in these results is the displacement of the sonic section, which is highlighted by the vertical dashed line and commented below. The sonic section is located in the geometrical throat (indicated by the red triangle) in the inviscid case, as predicted by the original compound flow theory (see, for example, \cite{bernstein1967compound}). The viscous simulation, without wall friction but with shear forces between the streams, shows an upstream shift towards the convergent section, as expected in the analysis of Section \ref{sec:choking}. Conversely, wall friction is dominant in the final simulation, overcoming the effects of the shear force and ultimately pushing the sonic section in the divergent portion of the nozzle.
		
	Note that the primary Mach number is slightly overestimated in all cases. This is attributed to the weak shock train at the inlet (see figures \ref{fig:comparison_f_corr}, \ref{fig:comparison_f_calib} and \ref{fig:2D_CFD}), which introduces a pressure loss in the primary stream.
	
\begin{figure}
		\center
		\begin{subfigure}[t]{0.48\linewidth}
			\includegraphics[width=0.99\linewidth]{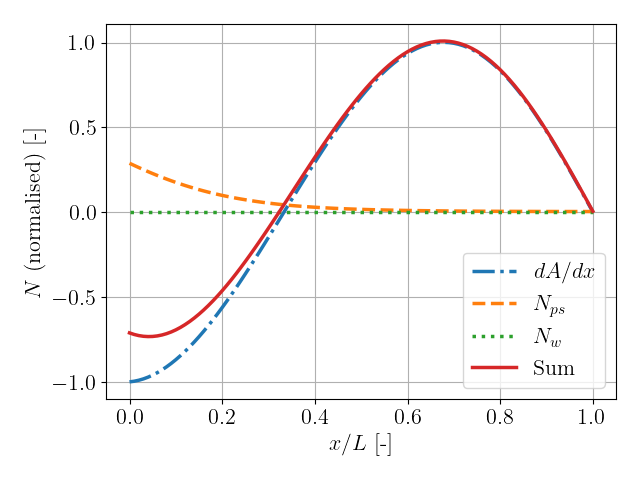}
			\subcaption{Slip, sonic at $x/L = 0.323$.}
			\label{fig:numerator_slip}
		\end{subfigure}
		\begin{subfigure}[t]{0.48\linewidth}
			\includegraphics[width=0.99\linewidth]{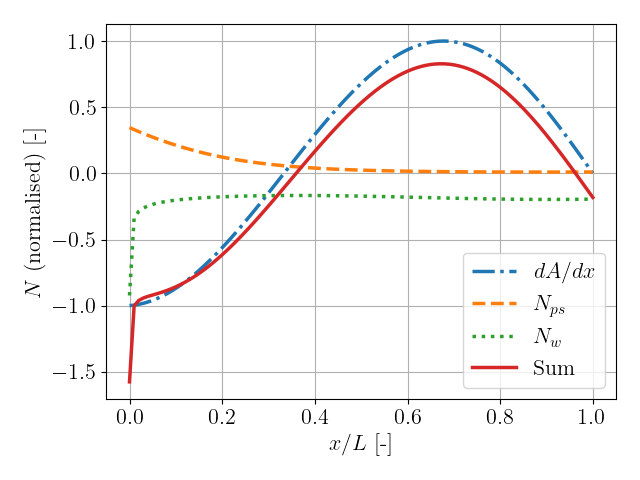}
			\subcaption{No slip, sonic at $x/L = 0.359$.}
			\label{fig:numerator_noslip}
		\end{subfigure}
		\caption{Components of the numerator in equation \eqref{eq:p_compound_2stream}, normalised by the maximal area gradient. The positive inter-stream friction term pushes the zero-crossing of the sum upstream, while the negative wall friction term pushes it downstream.}
		\label{fig:numerator}
	\end{figure}

	The 1D predictions allow for computing each term in the equation for the static pressure gradient (cf. equation \eqref{eq:p_compound_2stream}) separately. The evolution of the components in the numerator is shown in figure \ref{fig:numerator} for the slip and no-slip cases. The inviscid case is trivial since the gradient of the cross-section is the only remaining term. The terms are normalised with the maximal gradient of the cross-section, as this is a fixed quantity for the three simulations. The evolution of the two friction terms supports the theoretical considerations in Section \ref{sec:choking}, the first (inter-stream friction) always being positive and the second (wall friction) being negative. As a result, the zero-crossing of the numerator moves up or downstream. The gradient of the cross-section balances the effect of the forces in either direction by moving the sonic section in the convergent or the divergent section. Note that the inter-stream friction tends to decrease from left to right as the difference in Mach number decreases, while friction tends to increase as the flow accelerates through the nozzle.
		
	\subsection{Flow blockage mechanism} \label{sec:results_blockage}
	As observed in figure \ref{fig:1D_nozzle}, the secondary Mach number is below unity in the compound-sonic section if the primary stream is supersonic. It reaches unity further downstream, where the compound flow continues to expand in the diverging section of the nozzle. It can thus be stated that the \emph{compound} flow is sonic in the first section and that the \emph{secondary} stream is sonic, separately, in a section further downstream. Both sonic sections are plausible explanations for the blockage of the secondary mass flow rate, e.g., in ejectors, these choking mechanisms are referred to as `compound choking' and `Fabri choking' (see \cite{LAMBERTS2018_compound}). 
	There is currently no consensus on which of these mechanisms limits the secondary mass flow rate.
	
	The excellent match between the post-processed RANS simulations and the 1D predictions of the compound flow theory suggests that the compound choking condition is the more general mechanism. Furthermore, the secondary Mach number reaches unity necessarily \emph{downstream} of the compound-sonic section since a combination of a sonic and a supersonic stream is necessarily compound-supersonic (cf. equation \eqref{eq:beta}). This suggests that cases identified as `Fabri choking' also feature compound choking further upstream (see \cite{hoge1965choked}). Moreover, \cite{LAMBERTS2018_compound} and \cite{kracik2023effect} showed that the static pressure required to reach compound choking is higher than the one required to reach $\Ma_s = 1$. This implies that the compound condition is less restrictive (the flow needs to expand less to reach the compound condition), but still sufficient to choke the flow. As a definitive proof of the generality of the compound choking condition to explain the mass flow limitation (compared to Fabri choking), a compound choked flow is demonstrated with a fully subsonic secondary stream.
	
	To lower the secondary Mach number below unity, the divergent section of the nozzle presented in Section \ref{sec:numerics} is replaced by a constant cross-section having the same diameter as the throat. The first simulation was carried out on a nozzle of the same length, while the second is cut shorter at $x/L=0.4$ (so downstream of the throat at $x/L=0.333$). The settings for the mesh, the boundary conditions and the numerical solver remain identical to the previous simulations. The absence of a divergent section would severely impact the viscous case with the no-slip condition at the wall since the compound sonic section is located in the divergent section. The resulting flow field forms a Fanno flow, which becomes nearly sonic at the outlet of the constant-area pipe. Therefore, the viscous case with a slip condition at the wall is retained, which becomes compound-sonic upstream of the throat regardless of the geometry downstream.
	
\begin{figure}
		\center
		\begin{subfigure}[t]{0.48\linewidth}
			\includegraphics[width=0.99\linewidth]{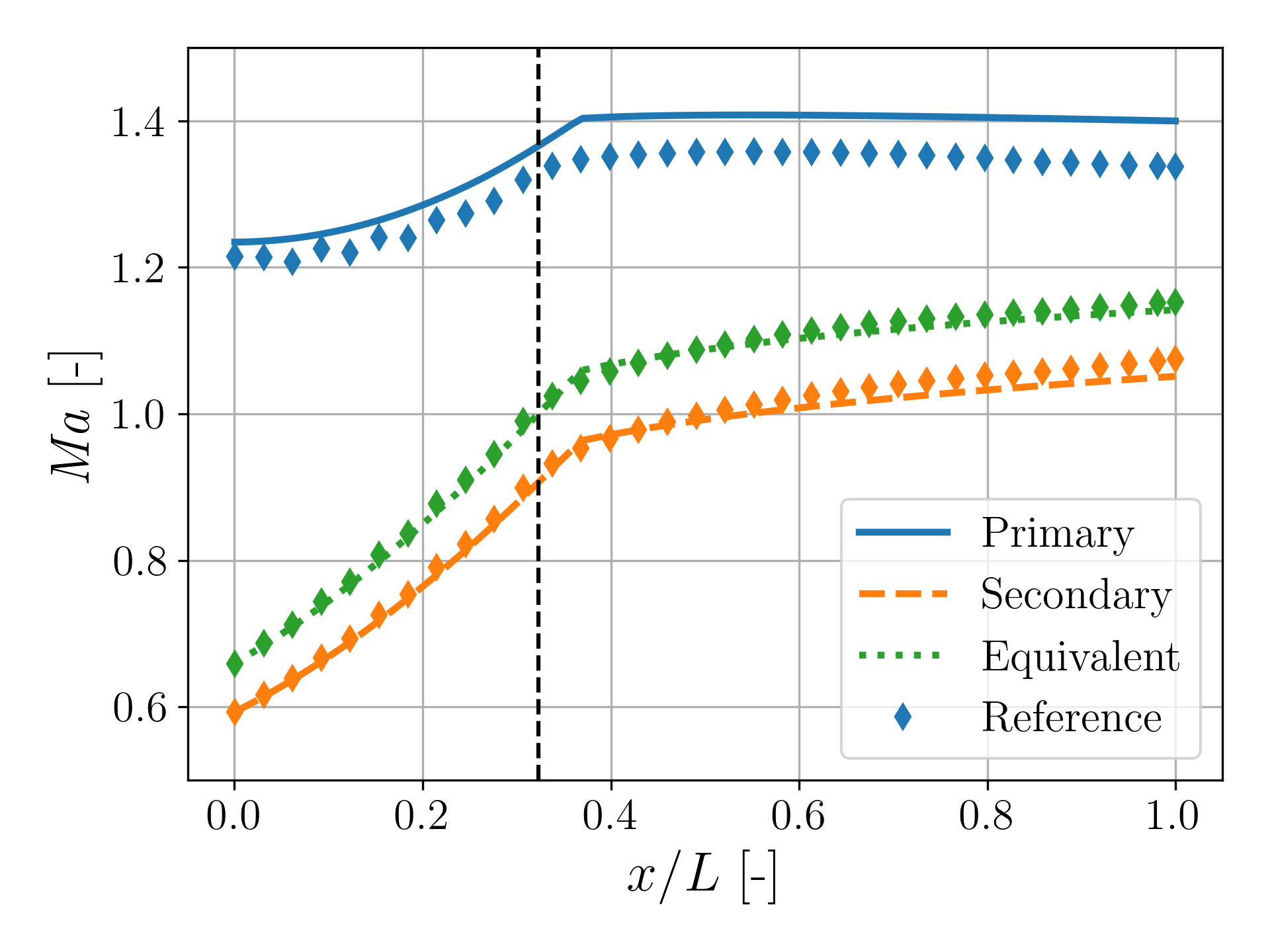}
			\subcaption{Case 2, convergent-constant configuration.}
			\label{fig:constant_area_Mach}
		\end{subfigure}
		\begin{subfigure}[t]{0.48\linewidth}
			\includegraphics[width=0.99\linewidth]{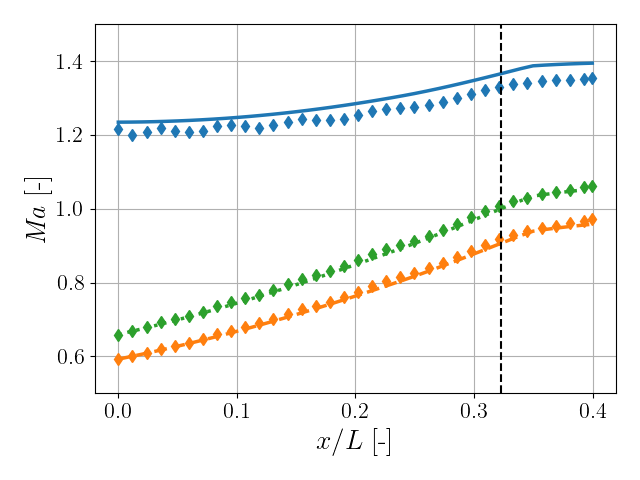}
			\subcaption{Case 2, cut configuration.}
			\label{fig:cut_area_Mach}
		\end{subfigure}
		\caption{Evolution of the Mach number for the convergent-constant and cut nozzle with friction between the streams and a slip condition at the wall (see figure \ref{fig:1D_nozzle_slip}). The results coincide in the convergent section as the flow is choked in all cases. The secondary Mach number does not reach unity in the shorter nozzle.}
		\label{fig:constant_area}
	\end{figure}

	Figure \ref{fig:constant_area} shows the evolution of the post-processed Mach number of the convergent-constant and cut (still convergent-constant) configurations. The compound choking condition is reached at $x/L = 0.323$ in both cases, and the secondary mass flow rates are identical to 0.0622 kg/s (since the geometry and the boundary conditions at the inlet are also identical). The flow expands less in the constant area section compared to the convergent-divergent case, but the secondary Mach number still reaches unity at $x/L = 0.56$. However, the secondary cross-section does not reach a minimum in this section, suggesting that the secondary stream does not choke separately. This is confirmed by the simulation with the shorter nozzle, which coincides with the results of the complete nozzle. Despite the low pressure imposed as a boundary condition at the outlet, the secondary flow does not accelerate beyond $\Ma_s = 0.97$ (both in the RANS simulation and in the 1D model). Hence, no larger secondary mass flow rate can be obtained, and the compound flow is choked with a fully subsonic secondary stream. These results suggest that compound choking is the more general choking condition in ejectors and, more generally, for parallel compressible streams with different stagnation pressures.
		
	\vspace{-4mm}
	\section{Conclusion} \label{sec:conclusion}
	The isentropic compound flow theory has been extended with friction forces between the streams and at the wall. The resulting 1D model captures the upstream and downstream displacement of the sonic section observed by \cite{LAMBERTS2018_compound} and \cite{kracik2023effect}. Moreover, the 1D predictions have been validated numerically to 2D axisymmetric RANS simulations with appropriate boundary conditions to include or eliminate the effect of friction and a parametric study has demonstrated that the model generalises well across different operating conditions and different geometries. An extension of this work to the modelling of ejectors is foreseen in the future.
	
	To the authors' knowledge, our work is the first to propose an explanation for choking of a compound flow \emph{upstream} of the geometrical throat.
	The Taylor expansion of the equation for the static pressure gradient allows for circumventing its singularity, avoiding numerical instabilities and increasing the accuracy of the numerical predictions. Analytical expressions have been presented for all required derivatives for a second-order approximation, with a quasi-third-order approximation obtained with finite differences. These expressions and the governing equations have first been derived in a general way in terms of a force $F_i$, facilitating future developments such as a compound flow with more than two streams or a planar geometry instead of the axisymmetrical one considered in this work. Another open path includes heat exchange and different total temperatures at the inlet, which may affect the shifting mechanism due to friction between the streams, as discussed in Section \ref{sec:choking}. In particular, a high primary inlet temperature may invert the effect of inter-stream friction. Finally, we believe that the developed model not only provides more insight into the choking of compound flows but also opens perspectives in the reduced-order modelling of ejectors.
		
	\vspace{3mm}
	\textbf{Acknowledgements} J. Van den Berghe is supported by F.R.S.-FNRS FRIA grant number 47455. Declaration of Interests. The authors report no conflict of interest.
	\appendix
	\section{First order approximation of the static pressure gradient}\label{sec:dp_dx_first_order}
The static pressure gradient in the sonic section has been calculated in Section \ref{sec:choking} using de l'Hôpital's rule (see \eqref{eq:p_delhopital}). This operation requires the first derivatives of the numerator $N$ and the compound choking indicator $\beta$. This section provides analytical expressions for these derivatives and those of the forces ($dF_i/dx$), ultimately leading to a quadratic equation for the static pressure gradient in the sonic section.

\vspace{-3mm}

\subsection{Friction forces}
As discussed in Section \ref{sec:choking}, the derivatives of the numerator and $\beta$ give rise to derivatives of Mach numbers $\Ma_i$, cross-section $A_i$, forces $F_i$ and the static pressure $p$. These can be substituted by the governing equations \eqref{eq:pt_i}, \eqref{eq:Tt_i}, \eqref{eq:A_i}, resulting in expressions that are linear with respect to the static pressure gradient. This is also true for the friction forces. The first derivative of the wall friction force is given by the following equation:
\begin{equation}
    \dfrac{d F_w}{dx} = c_{Fw0} \bigg(\dfrac{1}{p}\dfrac{dp}{dx}\bigg) + c_{Fw1}\,, \label{eq:dFw_dx}
\end{equation}
where
\begin{align}
    c_{Fw0} =& f_w \sqrt{\pi} \sqrt{A}  p \left(  \Ma_s^2-2\right)\,,\\
    c_{Fw1} =& \dfrac{\sqrt{\pi}\left(f_w \gamma A_s \Ma_s^2 p \dfrac{dA}{d x} +\left( 2 f_w \gamma-2 f_w\right)  A F_s \Ma_s^2+4 f_w A F_s\right)}{2 \sqrt{A} A_s}\,.
\end{align}
The inter-stream friction force explicitly depends on densities, velocities and the primary cross-section (see \eqref{eq:stream_friction}). The vector $\bm{q} = [\rho_p, \rho_s, u_p, u_s, A_p, A_s, \Ma_p^2, \Ma_s^2]$ is introduced to ease the notation. The first derivative of the friction between the streams is given by the following equation:
\begin{equation}
    \dfrac{d F_{ps}}{dx} = \dfrac{\partial F_{ps}}{\partial \bm{q}} \cdot \dfrac{d\bm{q}}{dx}\,, \label{eq:dFps_dx_start}
\end{equation}
where the dot operator indicates an inner product. The equations below are derived for the case where $u_p > u_s$ to avoid the absolute value operator in \eqref{eq:stream_friction}. The derivatives in the case $u_p < u_s$ have the same magnitude, but only differ in the sign. It thus suffices to add a minus sign in the equation above. The partial derivatives of $F_{ps}$ with respect to $\bm{q}$ are given by:
\begin{equation}\label{eq:dFps_dq}
    \dfrac{\partial F_{ps}}{\partial \bm{q}} =
    \begin{pmatrix}
    \dfrac{f_{ps} \sqrt{\pi} \sqrt{A_p} {{\left( u_p-u_s\right) }^{2}}}{2}\\
    \dfrac{f_{ps} \sqrt{\pi} \sqrt{A_p} {{\left( u_p-u_s\right) }^{2}}}{2}\\
    f_{ps} \sqrt{\pi} \sqrt{A_p} \left( \rho_s+\rho_p\right)  \left( u_p-u_s\right) \\
    -f_{ps} \sqrt{\pi} \sqrt{A_p} \left( \rho_s+\rho_p\right)  \left( u_p-u_s\right) \\
    \dfrac{f_{ps} \sqrt{\pi} \left( \rho_s+\rho_p\right)  {{\left( u_p-u_s\right) }^{2}}}{4 \sqrt{A_p}}\\
    0\\
    0\\
    0
    \end{pmatrix}\,.
\end{equation}
Note that the last three quantities of $\bm{q}$ do not appear directly in the definition of $F_{ps}$. These are included in $\bm{q}$ because they appear in the second derivatives (see Section \ref{sec:dp_dx_second_order}). The derivative $d\bm{q}/dx$ follows from the governing equations (see Section \ref{sec:derivation_governing} and \cite{shapiro1953dynamics}):
\begin{equation}\label{eq:dq_dx}
    \dfrac{d\bm{q}}{dx} = \bm{c}_{q0} \left(\dfrac{1}{p}\dfrac{dp}{dx}\right) + \bm{c}_{q1p} F_p + \bm{c}_{q1s} F_s\,,
\end{equation}
where $\bm{c}_{q0}$, $\bm{c}_{q1p}$ and $\bm{c}_{q1s}$ are vectors of the same length as $\bm{q}$, and themselves functions of $\bm{q}$:
\begin{equation}\label{eq:cq0}
    \bm{c}_{q0} =
    \begin{pmatrix}
    \dfrac{\rho_p}{\gamma}\\
    \dfrac{\rho_s}{\gamma}\\
    -\dfrac{u_p}{\gamma \Ma_p^2} \\
    -\dfrac{u_s}{\gamma \Ma_s^2} \\
    \dfrac{A_p}{\gamma} \left( \dfrac{1}{\Ma_p^2}-1\right)\\
    \dfrac{A_s}{\gamma} \left( \dfrac{1}{\Ma_s^2}-1\right)\\
    -\dfrac{2}{\gamma} \left( \dfrac{\left( \gamma-1\right)}{2} \Ma_p^2 +1\right)\\
    -\dfrac{2}{\gamma} \left( \dfrac{\left( \gamma-1\right)}{2} \Ma_s^2 +1\right)
    \end{pmatrix}\,,
\end{equation}
\begin{equation}\label{eq:cq1}
    \bm{c}_{q1p} =
    \begin{pmatrix}
    \dfrac{\left( \gamma-1\right)  \rho_p}{\gamma A_p p}\\
    0\\
    \dfrac{u_p}{\gamma A_p \Ma_p^2 p}\\
    0\\
    \dfrac{-\left( \dfrac{1}{\gamma \Ma_p^2}\right) -\dfrac{\gamma-1}{\gamma}}{p}\\
    0\\
    \dfrac{2 \left( \dfrac{\left( \gamma-1\right)  \Ma_p^2}{2}+1\right) }{\gamma A_p p}\\
    0
    \end{pmatrix}\,, \quad \mbox{and } \quad
    \bm{c}_{q1s} =
    \begin{pmatrix}
    0\\
    \dfrac{\left( \gamma-1\right)  \rho_s}{\gamma A_s p}\\
    0\\
    \dfrac{u_s}{\gamma A_s \Ma_s^2 p}\\
    0\\
    \dfrac{-\left( \dfrac{1}{\gamma \Ma_s^2}\right) -\dfrac{\gamma-1}{\gamma}}{p}\\
    0\\
    \dfrac{2 \left( \dfrac{\left( \gamma-1\right)  \Ma_s^2}{2}+1\right) }{\gamma A_s p}
    \end{pmatrix}\,.
\end{equation}
Combining \eqref{eq:dFps_dx_start} - \eqref{eq:cq1} leads to the following equation:
\begin{equation}
    \dfrac{d F_{ps}}{dx} = c_{Fps0} \bigg(\dfrac{1}{p}\dfrac{dp}{dx}\bigg) + c_{Fps1}\,, \label{eq:dFps_dx}
\end{equation}
where
\begin{equation}\label{eq:cFps0}
    c_{Fps0} = \dfrac{\partial F_{ps}}{\partial \bm{q}} \cdot \bm{c}_{q0}\,,
\end{equation}
and
\begin{equation}\label{eq:cFps1}
    c_{Fps1} = \dfrac{\partial F_{ps}}{\partial \bm{q}} \cdot \left(\bm{c}_{q1p} F_p + \bm{c}_{q1s} F_s\right)\,.
\end{equation}
The derivatives of the forces $F_i$ follow from \eqref{eq:dFw_dx}, \eqref{eq:dFps_dx} and their definition \eqref{eq:F_i}:
\begin{equation}
    \dfrac{dF_i}{dx} = c_{Fi0} \bigg(\dfrac{1}{p}\dfrac{dp}{dx}\bigg) + c_{Fi1}\,, \label{eq:dF_i_dx}
\end{equation}
where $i\in[p,s]$ and
\begin{align}
    &c_{Fp0} = -c_{Fps0}\,, \quad and \quad c_{Fs0} = c_{Fps0} - c_{Fw0}\,,\nonumber\\
    &c_{Fp1} = -c_{Fps1}\,, \quad and \quad c_{Fs1} = c_{Fps1} - c_{Fw1}\,, \label{eq:cFi}
\end{align}

\subsection{The numerator of the equation for the static pressure}
The numerator is differentiated in its general form \eqref{eq:p_compound} for any definition of the forces $F_i$ to facilitate potential modifications to the forces in future work:
\begin{equation}
    N =  \dfrac{dA}{dx} + \sum\limits_{i\in[p,s]}\left[\dfrac{1 + \left(\gamma  - 1\right)\Ma_i^2}{\gamma \Ma_i^2}\right] \dfrac{F_i}{p}\,.
\end{equation}
Introducing the concise notation for the second term:
\begin{equation}
    Z_i = \left[\dfrac{1 + \left(\gamma  - 1\right)\Ma_i^2}{\gamma \Ma_i^2}\right] \dfrac{F_i}{p}\,,\label{eq:Z_i}
\end{equation}
the gradient of the numerator gives rise to first derivatives of $Z_i$:
\begin{equation}\label{eq:dN_dx}
    \dfrac{dN}{dx} = \dfrac{d^2A}{dx^2} + \sum\limits_{i\in[p,s]} \dfrac{dZ_i}{dx}\,.
\end{equation}
These are given by:
\begin{equation}
    \dfrac{dZ_i}{dx} = c_{Zi0} \left( \dfrac{dF_i}{d x}\right) + c_{Zi1} \bigg(\dfrac{1}{p}\dfrac{dp}{dx}\bigg) + c_{Zi2}\,, \label{eq:dZi_dx}
\end{equation}
where
\begin{align}
    c_{Zi0} = &\dfrac{1}{p}\left(\dfrac{1 + \left(\gamma  - 1\right)\Ma_i^2}{\gamma \Ma_i^2}\right)\,,\label{eq:cZi0}\\
    c_{Zi1} = &\dfrac{\left(\left(\gamma-\gamma^2\right)\Ma_i^4 - \Ma_i^2 +2\right)F_i}{\gamma^2 p \Ma_i^4}\,,\label{eq:cZi1}\\
    c_{Zi2} = &-\dfrac{2 F_i^2 \left( \dfrac{\left( \gamma-1\right)  \Ma_i^2}{2}+1\right) }{\gamma^2 p^2 A_i \Ma_i^4}\,. \label{eq:cZi2}
\end{align}
The combination of \eqref{eq:dN_dx}, \eqref{eq:dZi_dx} and \eqref{eq:dF_i_dx} yields the final expression for the first derivative of the numerator:
\begin{equation}\label{eq:dN_dx_final}
    \dfrac{dN}{dx} = c_{N0} \bigg(\dfrac{1}{p}\dfrac{dp}{dx}\bigg)
    + c_{N1}\,,
\end{equation}
where
\begin{align}
    c_{N0} &= \sum\limits_{i\in[p,s]} \left(c_{Zi0} \, c_{Fi0} + c_{Zi1}\right)\,,\\
    c_{N1} &= \dfrac{d^2A}{dx^2} + \sum\limits_{i\in[p,s]} \left(c_{Zi0} \, c_{Fi1} + c_{Zi2}\right)\,,
\end{align}
which require \eqref{eq:cFi}, \eqref{eq:cZi0}, \eqref{eq:cZi1} and \eqref{eq:cZi2}.

\subsection{The compound choking indicator}
The compound choking indicator can be written as a sum of components $\beta_i$ (cf. \eqref{eq:beta}):
\begin{equation}
    \beta = \sum\limits_{i\in[p,s]} \beta_i = \sum\limits_{i\in[p,s]} A_i\dfrac{1-\Ma_i^2}{\gamma \Ma_i^2}.
\end{equation}
The first derivative of $\beta$ is given by the following equation:
\begin{equation} \label{eq:dbeta_dx_final}
    \dfrac{d\beta}{dx} = c_{\beta 0} \bigg(\dfrac{1}{p}\dfrac{dp}{dx}\bigg) + c_{\beta 1},
\end{equation}
where
\begin{align}
    c_{\beta 0} = & \sum\limits_{i\in[p,s]} \dfrac{A_i\left({\Ma_i^4}+\left( \gamma-3\right)  \Ma_i^2+3 \right)  }{\gamma^2 {\Ma_i^4}}\,,\label{eq:cbeta0}\\
    c_{\beta 1} = & \sum\limits_{i\in[p,s]} \dfrac{\left( \gamma-1\right)  {\Ma_i^4}+\left( 3-2 \gamma\right)  \Ma_i^2-3 }{\gamma^2  {\Ma_i^4}}\left(\dfrac{F_i}{p}\right)\,.\label{eq:cbeta1}
\end{align}

\subsection{Static pressure gradient}
Using de l'Hôpital's rule on \eqref{eq:p_compound} leads to \eqref{eq:p_quadratic}, which is equivalent to the following quadratic equation:
\begin{equation}\label{eq:app_p_quadratic}
     c_{\beta 0}^*\left(\dfrac{1}{p^*} \dfrac{dp}{dx}^*\right)^2 +  \left(c_{\beta1}^* - c_{N0}^*\right)\left(\dfrac{1}{p^*} \dfrac{dp}{dx}^*\right)
     -c_{N1}^* = 0\,.
\end{equation}
The stars indicate that the variables are evaluated at the sonic section. The two roots correspond to a subsonic or supersonic solution downstream of the sonic section. The first order approximation in the vicinity of the sonic section consists of the same constant gradient in the throat given by the equation above (see figure \ref{fig:dpdx}).

\section{Second order approximation of the static pressure gradient}\label{sec:dp_dx_second_order}
The second order approximation requires the first and the second derivatives of the numerator $N$ and $\beta$ (see \eqref{eq:p_taylor}). These give rise to second derivatives of the quantities encountered in Appendix \ref{sec:dp_dx_first_order}. In this section, the first derivatives are not expanded since they can be calculated directly using the expressions in Appendix \ref{sec:dp_dx_first_order}. 


\subsection{Mach number}
\begin{equation}
    \dfrac{d^2\Ma_i^2}{dx^2} = c_{\Ma_i^20} \left(  \dfrac{d}{d x} \bigg(\dfrac{1}{p}\dfrac{dp}{dx}\bigg) \right)  + c_{\Ma_i^21}\,,
\end{equation}
where
\begin{align}
    c_{\Ma_i^20} = &- \dfrac{2}{\gamma} \left( \dfrac{\left( \gamma-1\right)  \Ma_i^2}{2}+1\right)\,,\\
    c_{\Ma_i^21} =
    &-\dfrac{\left( \gamma-1\right)}{\gamma}  \bigg(\dfrac{1}{p}\dfrac{dp}{dx}\bigg) \left( \dfrac{d \Ma_i^2}{d x}\right)  +\dfrac{\left( \gamma-1\right)}{\gamma}  \dfrac{F_i}{p A_i} \left( \dfrac{d \Ma_i^2}{d x}\right) \nonumber\\
    &+\dfrac{2}{\gamma p A_i} \left( \dfrac{\left( \gamma-1\right)  \Ma_i^2}{2}+1\right)  \left( \dfrac{dF_i}{d x}\right)  - \dfrac{2 F_i}{\gamma p A_i} \left( \dfrac{\left( \gamma-1\right)  \Ma_i^2}{2}+1\right)  \left( \dfrac{1}{p}\dfrac{dp}{d x}\right)\,,
\end{align}
where the derivatives $d\Ma_i^2/dx$ are the last two components of the vector $d\bm{q}/dx$ in \eqref{eq:dq_dx}.

\subsection{Friction forces}
The second derivative of the wall friction force is given by the following equation:
\begin{equation}\label{eq:d2Fw_dx2}
    \dfrac{d^2 F_w}{dx^2} = c_{Fw^20} \left( \dfrac{d}{d x} \bigg(\dfrac{1}{p}\dfrac{dp}{dx}\bigg)\right) + c_{Fw^21}\,,
\end{equation}
where
\begin{align}
    c_{Fw^20} = &f_w \sqrt{\pi} \sqrt{A} p \left(\Ma_s^2 - 2\right)\,,\\
    c_{Fw^21} = &f_w \gamma \sqrt{\pi} \sqrt{A} p c_{\Ma_s^21} +\dfrac{f_w \gamma \sqrt{\pi} p }{\sqrt{A}} \left( \dfrac{d A}{d x}\right) \left( \dfrac{d \Ma_s^2}{d x}\right) +2 f_w \gamma \sqrt{\pi} \sqrt{A} p \bigg(\dfrac{1}{p}\dfrac{dp}{dx}\bigg)  \left( \dfrac{d \Ma_s^2}{d x}\right) \nonumber\\
    &+\dfrac{f_w \gamma \sqrt{\pi} \Ma_s^2 p}{2 \sqrt{A}} \left( \dfrac{d^2 A}{dx^2}\right) -\dfrac{f_w \gamma \sqrt{\pi} \Ma_s^2 p}{4 {{A}^{\frac{3}{2}}}} \left( \dfrac{d A}{d x}\right)^2 \nonumber\\
    &+\dfrac{f_w \gamma \sqrt{\pi} \Ma_s^2 p}{\sqrt{A}} \bigg(\dfrac{1}{p}\dfrac{dp}{dx}\bigg) \left( \dfrac{d A}{d x}\right) +f_w \gamma \sqrt{\pi} \sqrt{A} \Ma_s^2 {{\bigg(\dfrac{1}{p}\dfrac{dp}{dx}\bigg)}^{2}} p\,.
\end{align}

The second derivative of the inter-stream friction force follows from \eqref{eq:dFps_dx} (with the same comment regarding the sign in case $u_p < u_s$):
\begin{equation}
    \dfrac{d^2 F_{ps}}{dx^2} = c_{Fps0} \left(\dfrac{d}{dx}\bigg(\dfrac{1}{p}\dfrac{dp}{dx}\bigg)\right) + \dfrac{d c_{Fps0}}{dx} \bigg(\dfrac{1}{p}\dfrac{dp}{dx}\bigg) + \dfrac{d c_{Fps1}}{dx}\,.
\end{equation}
The derivatives of the coefficients $c_{Fps0}$ and $c_{Fps1}$ follow from \eqref{eq:cFps0} and \eqref{eq:cFps1}:
\begin{align}
    \label{eq:dcFps0_dx}
    \dfrac{dc_{Fps0}}{dx} = &\left(\dfrac{\partial^2 F_{ps}}{\partial \bm{q}^2} \dfrac{d\bm{q}}{dx}\right) \cdot \bm{c}_{q0} + \dfrac{\partial F_{ps}}{\partial \bm{q}} \cdot \dfrac{d\bm{c}_{q0}}{dx}\,, \\
    \dfrac{dc_{Fps1}}{dx} = &\left(\dfrac{\partial^2 F_{ps}}{\partial \bm{q}^2} \dfrac{d\bm{q}}{dx}\right) \cdot \left(\bm{c}_{q1p} F_p + \bm{c}_{q1s} F_s\right) \nonumber\\
    \label{eq:dcFps1_dx}
    &+ \dfrac{\partial F_{ps}}{\partial \bm{q}} \cdot \left(\dfrac{d\bm{c}_{q1p}}{dx} F_p + \bm{c}_{q1p}\dfrac{dF_p}{dx} + \dfrac{d\bm{c}_{q1s}}{dx} F_s + \bm{c}_{q1s} \dfrac{dF_s}{dx} \right)\,.
\end{align}
The first derivatives $\partial F_{ps}/\partial \bm{q}$, $d\bm{q}/dx$, $dF_p/dx$ and $dF_s/dx$ are given by \eqref{eq:dFps_dq}, \eqref{eq:dq_dx} and \eqref{eq:dF_i_dx}. The second derivatives of the friction force $F_{ps}$ with respect to $\bm{q}$ are given by the following matrix:
\begin{equation}
    \dfrac{\partial^2 F_{ps}}{\partial \bm{q}^2} = 
    f_{ps} \sqrt{\pi} \sqrt{A_p} \left(u_p-u_s\right) \bm{M}\,,
\end{equation}
where
\begin{equation}
    \bm{M} = \begin{pmatrix}
    0 & 0 & 1 & -1 & \dfrac{u_p-u_s}{4 A_p} & 0 & 0 & 0\\
    0 & 0 & 1 & -1 & \dfrac{ u_p-u_s}{4 A_p} & 0 & 0 & 0 \\
    1 & 1 & \dfrac{\rho_p+\rho_s}{u_p-u_s} & -\dfrac{\rho_p+\rho_s}{u_p-u_s} & \dfrac{\rho_p+\rho_s}{2 A_p} & 0 & 0 & 0 \\
    -1 & -1 & -\dfrac{\rho_p+\rho_s}{u_p-u_s} & \dfrac{\rho_p+\rho_s}{u_p-u_s} & - \dfrac{\rho_p+\rho_s}{2 A_p} & 0 & 0 & 0 \\
    \dfrac{u_p-u_s}{4 A_p} & \dfrac{u_p-u_s}{4 A_p} & \dfrac{\rho_p+\rho_s}{2 A_p} & -\dfrac{\rho_p+\rho_s}{2 A_p} & - \dfrac{ \left( \rho_p+\rho_s\right)  \left( u_p-u_s\right)}{8 {{A_p}^2}} & 0 & 0 & 0 \\
    0 & 0 & 0 & 0 & 0 & 0 & 0 & 0 \\
    0 & 0 & 0 & 0 & 0 & 0 & 0 & 0 \\
    0 & 0 & 0 & 0 & 0 & 0 & 0 & 0 \\
    \end{pmatrix}\,.
\end{equation}
The derivative of $\bm{c}_{q0}$ follows from \eqref{eq:cq0}:
\begin{equation}
    \dfrac{d\bm{c}_{q0}}{dx} = \dfrac{\partial\bm{c}_{q0}}{\partial \bm{q}} \dfrac{d\bm{q}}{dx}\,,
\end{equation}
where
\begin{equation}
    \dfrac{\partial \bm{C}_{q0}}{\partial \bm{q}} =
    \begin{pmatrix}
    \dfrac{1}{\gamma} & 0 & 0 & 0 & 0 & 0 & 0 & 0 \\
    0 & \dfrac{1}{\gamma} & 0 & 0 & 0 & 0 & 0 & 0 \\
    0 & 0 & - \dfrac{1}{\gamma \Ma_p^2} & 0 & 0 & 0 & \dfrac{u_p}{\gamma \Ma_p^4} & 0 \\
    0 & 0 & 0 & - \dfrac{1}{\gamma \Ma_s^2} & 0 & 0 & 0 & \dfrac{u_s}{\gamma \Ma_s^4} \\
    0 & 0 & 0 & 0 & \dfrac{\dfrac{1}{\Ma_p^2}-1}{\gamma} & 0 & - \dfrac{A_p}{\gamma \Ma_p^4} & 0 \\
    0 & 0 & 0 & 0 & 0 & \dfrac{\dfrac{1}{\Ma_s^2}-1}{\gamma} & 0 & - \dfrac{A_s}{\gamma \Ma_s^4} \\
    0 & 0 & 0 & 0 & 0 & 0 & - \dfrac{\gamma-1}{\gamma} & 0 \\
    0 & 0 & 0 & 0 & 0 & 0 & 0 & - \dfrac{\gamma-1}{\gamma}
    \end{pmatrix}\,.
\end{equation}
The coefficients $\bm{c}_{q1p}$ and $\bm{c}_{q1s}$ explicitly depend on $\bm{q}$ and the static pressure $p$ (see \eqref{eq:cq1}), so their derivatives are given by the following equations:
\begin{align}
    \dfrac{d\bm{c}_{q1p}}{dx} = &\dfrac{\partial\bm{c}_{q1p}}{\partial \bm{q}} \dfrac{d\bm{q}}{dx} + \dfrac{\partial \bm{c}_{q1p}}{\partial p} \dfrac{dp}{dx}\,,\\
    \dfrac{d\bm{c}_{q1s}}{dx} = &\dfrac{\partial\bm{c}_{q1s}}{\partial \bm{q}} \dfrac{d\bm{q}}{dx} + \dfrac{\partial \bm{c}_{q1s}}{\partial p} \dfrac{dp}{dx} \,,
\end{align}
where
\begin{equation}
    \dfrac{\partial \bm{C}_{q1p}}{\partial \bm{q}} =
    \begin{pmatrix}
    \dfrac{\gamma-1}{\gamma A_p p} & 0 & 0 & 0 & -\dfrac{\left( \gamma-1\right)  \rho_p}{\gamma {{A_p}^{2}} p} & 0 & 0 & 0 \\
    0 & 0 & 0 & 0 & 0 & 0 & 0 & 0 \\
    0 & 0 & \dfrac{1}{\gamma A_p \Ma_p^2 p} & 0 & -\dfrac{u_p}{\gamma {{A_p}^{2}} \Ma_p^2 p} & 0 & -\dfrac{u_p}{\gamma A_p \Ma_p^4 p} & 0 \\
    0 & 0 & 0 & 0 & 0 & 0 & 0 & 0 \\
    0 & 0 & 0 & 0 & 0 & 0 & \dfrac{1}{\gamma \Ma_p^4 p} & 0 \\
    0 & 0 & 0 & 0 & 0 & 0 & 0 & 0 \\
    0 & 0 & 0 & 0 & -\dfrac{2 \left( \dfrac{\left( \gamma-1\right)  \Ma_p^2}{2}+1\right) }{\gamma {{A_p}^{2}} p} & 0 & \dfrac{\gamma-1}{\gamma A_p p} & 0 \\
    0 & 0 & 0 & 0 & 0 & 0 & 0 & 0
    \end{pmatrix}\,,
\end{equation}
\begin{equation}
    \dfrac{\partial \bm{C}_{q1s}}{\partial \bm{q}} =
    \begin{pmatrix}
    0 & 0 & 0 & 0 & 0 & 0 & 0 & 0 \\
    0 & \dfrac{\gamma-1}{\gamma A_s p} & 0 & 0 & 0 & -\dfrac{\left( \gamma-1\right)  \rho_s}{\gamma A_s^2 p} & 0 & 0 \\
    0 & 0 & 0 & 0 & 0 & 0 & 0 & 0 \\
    0 & 0 & 0 & \dfrac{1}{\gamma A_s \Ma_s^2 p} & 0 & -\dfrac{u_s}{\gamma A_s^2 \Ma_s^2 p} & 0 & -\dfrac{u_s}{\gamma A_s \Ma_s^4 p} \\
    0 & 0 & 0 & 0 & 0 & 0 & 0 & 0 \\
    0 & 0 & 0 & 0 & 0 & 0 & 0 & \dfrac{1}{\gamma \Ma_s^4 p} \\
    0 & 0 & 0 & 0 & 0 & 0 & 0 & 0 \\
    0 & 0 & 0 & 0 & 0 & -\dfrac{2 \left( \dfrac{\left( \gamma-1\right)  \Ma_s^2}{2}+1\right) }{\gamma A_s^2 p} & 0 & \dfrac{\gamma-1}{\gamma A_s p}
    \end{pmatrix}\,,
\end{equation}
\begin{equation}
    \dfrac{\partial \bm{C}_{q1p}}{\partial p} =
    \begin{pmatrix}
    -\dfrac{\left( \gamma-1\right)  \rho_p}{\gamma A_p {p^2}}\\
    0\\
    -\dfrac{u_p}{\gamma A_p \Ma_p^2 {p^2}}\\
    0\\
    \dfrac{\left( \dfrac{1}{\gamma \Ma_p^2}\right) +\dfrac{\gamma-1}{\gamma}}{{p^2}}\\
    0\\
    -\dfrac{2 \left( \dfrac{\left( \gamma-1\right)  \Ma_p^2}{2}+1\right) }{\gamma A_p {p^2}}\\
    0
    \end{pmatrix}\,, \quad \mbox{and } \quad
    \dfrac{\partial \bm{C}_{q1s}}{\partial p} =
    \begin{pmatrix}
    0\\
    -\dfrac{\left( \gamma-1\right)  \rho_s}{\gamma A_s {p^2}}\\
    0\\
    -\dfrac{u_s}{\gamma A_s \Ma_s^2 {p^2}}\\
    0\\
    \dfrac{\left( \dfrac{1}{\gamma \Ma_s^2}\right) + \dfrac{\gamma-1}{\gamma}}{{p^2}}\\
    0\\
    -\dfrac{2 \left( \dfrac{\left( \gamma-1\right)  \Ma_s^2}{2}+1\right) }{\gamma A_s {p^2}}
    \end{pmatrix}\,.
\end{equation}
The second derivative of the inter-stream friction is thus given by:
\begin{equation}\label{eq:d2Fps_dx2}
    \dfrac{d^2 F_{ps}}{dx^2} = c_{Fps^20} \left(\dfrac{d}{dx}\bigg(\dfrac{1}{p}\dfrac{dp}{dx}\bigg)\right) + c_{Fps^21}\,,
\end{equation}
where
\begin{equation}
    c_{Fps^20} = c_{Fps0}\, \quad \mbox{and } \quad c_{Fps^21} = \dfrac{d c_{Fps0}}{dx} \bigg(\dfrac{1}{p}\dfrac{dp}{dx}\bigg) + \dfrac{d c_{Fps1}}{dx}\,
\end{equation}
are found with \eqref{eq:cFps0}, \eqref{eq:dcFps0_dx}, \eqref{eq:dcFps1_dx} and the local pressure gradient.
The second derivatives of the forces $F_i$ follow from \eqref{eq:d2Fw_dx2}, \eqref{eq:d2Fps_dx2} and their definition \eqref{eq:F_i}:
\begin{equation}
    \dfrac{d^2F_i}{dx^2} = c_{Fi^20} \left(\dfrac{d}{dx}\bigg(\dfrac{1}{p}\dfrac{dp}{dx}\bigg)\right) + c_{Fi^21}\,, \label{eq:d2Fi_dx2}
\end{equation}
where $i\in[p,s]$ and
\begin{align}
    &c_{Fp^20} = -c_{Fps^20}\,, \quad \mbox{and } \quad c_{Fs^20} = c_{Fps^20} - c_{Fw^20}\,,\nonumber\\
    &c_{Fp^21} = -c_{Fps^21}\,, \quad \mbox{and } \quad c_{Fs^21} = c_{Fps^21} - c_{Fw^21}\,, \label{eq:cF2i}
\end{align}

\subsection{The numerator of the equation for the static pressure}
The second derivative of the numerator gives rise to the second derivatives of $Z_i$:
\begin{equation}\label{eq:d2N_dx2}
    \dfrac{d^2N}{dx^2} = \dfrac{d^3A}{dx^3} + \sum\limits_{i\in[p,s]} \dfrac{d^2Z_i}{dx^2}
\end{equation}
The second derivative of the force term $Z_i$ (cf. equation \eqref{eq:Z_i}) is given by the following equation:
\begin{align} \label{eq:d2Zi_dx2}
    \dfrac{d^2Z_i}{dx^2} = &c_{Zi^20} \left( \dfrac{d}{d x} \left(\dfrac{1}{p}\dfrac{dp}{dx}\right)\right) 
    +c_{Zi^21} \left(\dfrac{{d^2F_i}}{dx^2}\right)
    +c_{Zi^22} \left( \dfrac{dF_i}{d x}\right)\nonumber\\
    &+ c_{Zi^23} \left(\dfrac{1}{p}\dfrac{dp}{dx}\right)^{2}
    + c_{Zi^24} \left(\dfrac{1}{p}\dfrac{dp}{dx}\right) + C_{Z^25,i}\,,
\end{align}
where
\begin{align}
    \label{eq:cZi20}
    c_{Zi^20} = &-\dfrac{\left( \gamma^2-\gamma\right)  F_i \Ma_i^4+ F_i \Ma_i^2-2 F_i}{\gamma^2 p \Ma_i^4}\,,\\
    \label{eq:cZi21}
    c_{Zi^21} = &\dfrac{\left( \gamma-1\right)  \Ma_i^2+1}{\gamma p \Ma_i^2}\,,\\
    c_{Zi^22} = &-\dfrac{\left( 2 \gamma^2-2 \gamma\right)  \Ma_i^4 + 2 \Ma_i^2 - 4}{\gamma^2 p \Ma_i^4}\left(\dfrac{1}{p}\dfrac{dp}{dx}\right) - \dfrac{\left(3 \gamma-3 \right) F_i \Ma_i^2+6 F_i}{\gamma^2 p^2 A_i \Ma_i^4}\,,\\
    c_{Zi^23} = &-\dfrac{\left( \gamma^2-\gamma^3\right) F_i \Ma_i^6- F_i \Ma_i^4 +\left( 6 - 2\gamma\right)  F_i \Ma_i^2-8 F_i}{\gamma^3 p \Ma_i^6}\,,\\
    c_{Zi^24} = &-\dfrac{ \left(1-\gamma^2\right)F_i^2 \Ma_i^4+\left( 5 \gamma -9\right) F_i^2 \Ma_i^2+14 F_i^2}{\gamma^3 p^2 A_i \Ma_i^6}\,,\\
    c_{Zi^25} = &\dfrac{\left( 3 \gamma-3\right)  {{F_i}^{3}} \Ma_i^2+6 {{F_i}^{3}}}{\gamma^3 p^3 A_i^2 \Ma_i^6}\,.
\end{align}
Grouping the terms excluding the second derivatives:
\begin{equation}
    \label{eq:cZi225}
    c_{Zi^22-5} = c_{Zi^22} \left( \dfrac{dF_i}{d x}\right)
    + c_{Zi^23} \left(\dfrac{1}{p}\dfrac{dp}{dx}\right)^{2}
    + c_{Zi^24} \left(\dfrac{1}{p}\dfrac{dp}{dx}\right) + C_{Zi^25}\,.
\end{equation}

The combination of \eqref{eq:d2N_dx2}, \eqref{eq:d2Zi_dx2} and \eqref{eq:d2Fi_dx2} yields the final expression for the second derivative of the numerator:
\begin{equation}\label{eq:d2N_dx2_final}
    \dfrac{d^2N}{dx^2} = c_{N^20} \, \left(\dfrac{d}{dx}\bigg(\dfrac{1}{p}\dfrac{dp}{dx}\bigg)\right)
    + c_{N^21}\,,
\end{equation}
where
\begin{align}
    \label{eq:cN20}
    c_{N^20} &= \sum\limits_{i\in[p,s]} \left(c_{Zi^20} + c_{Zi^21} \, c_{Fi^20}\right)\,,\\
    \label{eq:cN21}
    c_{N^21} &= \dfrac{d^3A}{dx^3} + \sum\limits_{i\in[p,s]} \left(c_{Zi^21} \, c_{Fi^21} + c_{Zi^22-5}\right)\,,
\end{align}
which require \eqref{eq:cF2i}, \eqref{eq:cZi20}, \eqref{eq:cZi21} and \eqref{eq:cZi225}.

\subsection{The compound choking indicator}
The second derivative of $\beta_i$ is given by the following equation:
\begin{equation} \label{eq:d2beta_dx2}
    \dfrac{d^2\beta_i}{dx^2} = c_{\beta i^2 0}  \left( \dfrac{d}{d x} \left(\dfrac{1}{p}\dfrac{dp}{dx}\right)\right) 
    +c_{\beta i^2 1} \left( \dfrac{dF_i}{d x}\right) + c_{\beta i^2 2}\left(\dfrac{1}{p}\dfrac{dp}{dx}\right)^{2} + c_{\beta i^2 3} \left(\dfrac{1}{p}\dfrac{dp}{dx}\right)  + c_{\beta i^2 4} \,,
\end{equation}
where
\begin{align}
    \label{eq:cbetai20}
    c_{\beta i^2 0} = &\dfrac{A_i {\Ma_i^4}+\left( \gamma-3 \right) A_i \Ma_i^2+3 A_i}{\gamma^2  \Ma_i^4}\,,\\
    c_{\beta i^2 1} = &\dfrac{\left( \gamma-1\right)    {\Ma_i^4}+\left( 3 -2 \gamma\right)    {\Ma_i^2}-3 }{\gamma^2 p  \Ma_i^4}\,,\\
    c_{\beta i^2 2} = &\dfrac{- A_i \Ma_i^6 +\left( \gamma^2-5 \gamma+7\right)   A_i {\Ma_i^4}+\left( 9 \gamma-18\right)   A_i \Ma_i^2+15  A_i}{\gamma^3  \Ma_i^6}\,,\\
    c_{\beta i^2 3} = &\dfrac{\left( 1-\gamma^2\right)  F_i \Ma_i^6    -\left(  2 \gamma^2 -10 \gamma+10\right)  F_i \Ma_i^4+\left( 30-19 \gamma\right)  F_i \Ma_i^2-27 F_i}{\gamma^3 p \Ma_i^6}\,,\\
    c_{\beta i^2 4} = &\dfrac{\left( 2 {\gamma^2}-5 \gamma+3\right)  F_i^2 \Ma_i^4+\left( 10 \gamma-12\right)  F_i^2 \Ma_i^2+12 F_i^2}{\gamma^3 p^2 A_i \Ma_i^6}\,.
\end{align}
Grouping the terms excluding the second derivative of the pressure:
\begin{equation}\label{eq:cbetai214}
    c_{\beta i^2 1-4} = c_{\beta i^2 1} \left( \dfrac{dF_i}{d x}\right) + c_{\beta i^2 2}\left(\dfrac{1}{p}\dfrac{dp}{dx}\right)^{2} + c_{\beta i^2 3} \left(\dfrac{1}{p}\dfrac{dp}{dx}\right)  + c_{\beta i^2 4}\,,
\end{equation}
the second derivative of $\beta$ becomes:
\begin{equation}\label{eq:d2beta_dx2_final}
    \dfrac{d^2\beta}{dx^2} = c_{\beta^20} \, \left(\dfrac{d}{dx}\bigg(\dfrac{1}{p}\dfrac{dp}{dx}\bigg)\right)
    + c_{\beta^21}\,,
\end{equation}
where
\begin{align}
    \label{eq:cbeta20}
    c_{\beta^20} &= \sum\limits_{i\in[p,s]} c_{\beta i^20}\,,\\
    \label{eq:cbeta21}
    c_{\beta^21} &= \sum\limits_{i\in[p,s]} c_{\beta i^2 1-4}\,,
\end{align}
which require \eqref{eq:cbetai20} and \eqref{eq:cbetai214}.

\subsection{Static pressure}
The second derivative of the static pressure in the sonic section is given by:
\begin{equation}
    \dfrac{d}{d x} \left(\dfrac{1}{p}\dfrac{dp}{dx}\right) = \dfrac{d}{dx}\left(\dfrac{N}{\beta}\right) = \dfrac{\beta (dN/dx) - N (d\beta/dx)}{\beta^2}\,.
\end{equation}
The equation above can be evaluated using \eqref{eq:dN_dx_final} and \eqref{eq:dbeta_dx_final} if $\beta \neq 0$. Otherwise, de l'Hôpital's rule yields the following equation:
\begin{equation}
    \left(\dfrac{d}{d x} \left(\dfrac{1}{p}\dfrac{dp}{dx}\right)\right)^* = \dfrac{0}{0} = \dfrac{(d^2N/dx^2)^*}{2 (d\beta/dx)^*} - \dfrac{N^*}{\beta^*}\dfrac{(d^2\beta/dx^2)^*}{2(d\beta/dx)^*}\,,
\end{equation}
where $N^*/\beta^*$ equals the static pressure gradient in the sonic section (see \eqref{eq:p_delhopital} and \eqref{eq:app_p_quadratic}). The second derivatives $d^2N/dx^2$ and $d^2\beta/dx^2$ are given by \eqref{eq:d2N_dx2_final} and \eqref{eq:d2beta_dx2_final}, and are linear with respect to the second derivative of the static pressure:
\begin{align}
    \left(\dfrac{d}{d x} \left(\dfrac{1}{p}\dfrac{dp}{dx}\right)\right)^* = 
    &\dfrac{1}{2 (d\beta/dx)^*} \left(c_{N^20}^*  - \left(\dfrac{1}{p}\dfrac{dp}{dx}\right)^* c_{\beta^20}^* \right) \left(\dfrac{d}{dx}\bigg(\dfrac{1}{p}\dfrac{dp}{dx}\bigg)\right)^* \nonumber\\
    &+ \dfrac{1}{2 (d\beta/dx)^*} \left(c_{N^21}^* - \left(\dfrac{1}{p}\dfrac{dp}{dx}\right)^* c_{\beta^21}^*\right)\,.
\end{align}
The second derivative of the static pressure in the sonic section is finally given by:
\begin{equation}
    \left(\dfrac{d}{dx}\bigg(\dfrac{1}{p}\dfrac{dp}{dx}\bigg)\right)^* = \dfrac{- c_{N^21}^* + \left(\dfrac{1}{p}\dfrac{dp}{dx}\right)^* c_{\beta^21}^*}{c_{N^20}^*  - \left(\dfrac{1}{p}\dfrac{dp}{dx}\right)^* c_{\beta^20}^* - 2 \left(\dfrac{d\beta}{dx}\right)^*} \,,
\end{equation}
which requires \eqref{eq:dbeta_dx_final}, \eqref{eq:app_p_quadratic}, \eqref{eq:cN20}, \eqref{eq:cN21}, \eqref{eq:cbeta20} and \eqref{eq:cbeta21}. The second order approximation of the static pressure gradient in the sonic section follows from the Taylor expansion in equation \eqref{eq:p_taylor}, for which the first and second derivatives are given by \eqref{eq:dN_dx_final}, \eqref{eq:dbeta_dx_final}, \eqref{eq:d2N_dx2_final} and \eqref{eq:d2beta_dx2_final}. A quasi-third order approximation is obtained by computing the second derivatives on multiple grid points and applying finite differences to approximate the required third derivatives.

	\bibliographystyle{jfm}
	\bibliography{bibliography}
	
	
		

	

\end{document}